\documentclass[aoas]{imsart}
\RequirePackage{amsthm,amsmath,amsbsy,amsfonts}
\usepackage{bbm}
\usepackage{bm,graphicx,psfrag,epsf}
\usepackage{natbib}
\usepackage{url} 
\usepackage{placeins}
\usepackage{algorithm}
\usepackage{algpseudocode}
\usepackage{subfig} 
\RequirePackage[colorlinks,citecolor=blue,urlcolor=blue]{hyperref}

\startlocaldefs

\usepackage[usenames]{color}
\usepackage{xspace}

\definecolor{blue}{rgb}{0.2,0.5,0.7}
\definecolor{green}{rgb}{0.3,0.68,0.29}
\definecolor{purple}{rgb}{0.6,0.31,0.64}

\newtheorem{proposition}{Proposition}[section]
\ifpdf
\newcommand{\Sec}[1]{\hyperref[sec:#1]{Section~\ref*{sec:#1}}} 
\newcommand{\App}[1]{\hyperref[sec:#1]{Appendix~\ref*{sec:#1}}} 
\newcommand{\Eqn}[1]{\hyperref[eq:#1]{{\rm (\ref*{eq:#1})}}} 
\newcommand{\Part}[1]{\hyperref[part:#1]{(\ref*{part:#1})}} 
\newcommand{\Fig}[1]{\hyperref[fig:#1]{Figure~\ref*{fig:#1}}} 
\newcommand{\Tab}[1]{\hyperref[tab:#1]{Table~\ref*{tab:#1}}} 
\newcommand{\Thm}[1]{\hyperref[thm:#1]{Theorem~\ref*{thm:#1}}} 
\newcommand{\Lem}[1]{\hyperref[lem:#1]{Lemma~\ref*{lem:#1}}} 
\newcommand{\Prop}[1]{\hyperref[prop:#1]{Proposition~\ref*{prop:#1}}} 
\newcommand{\Cor}[1]{\hyperref[cor:#1]{Corollary~\ref*{cor:#1}}} 
\newcommand{\Def}[1]{\hyperref[def:#1]{Definition~\ref*{def:#1}}} 
\newcommand{\Alg}[1]{\hyperref[alg:#1]{Algorithm~\ref*{alg:#1}}} 
\newcommand{\Ex}[1]{\hyperref[ex:#1]{Example~\ref*{ex:#1}}} 
\newcommand{\As}[1]{\hyperref[as:#1]{Assumption~{\rm\ref*{as:#1}}}} 
\newcommand{\Reg}[1]{\hyperref[as:#1]{Condition~\ref*{reg:#1}}} 
\newcommand{\AlgLine}[2]{\hyperref[alg:#1]{line~\ref*{line:#2} of Algorithm~\ref*{alg:#1}}}
\newcommand{\AlgLines}[3]{\hyperref[alg:#1]{lines~\ref*{line:#2}--\ref*{line:#3} of Algorithm~\ref*{alg:#1}}}
\else
\newcommand{\Sec}[1]{{Section~\ref{sec:#1}}} 
\newcommand{\App}[1]{{Appendix~\ref{sec:#1}}} 
\newcommand{\Eqn}[1]{{(\ref{eq:#1})}} 
\newcommand{\Part}[1]{{(\ref{part:#1})}} 
\newcommand{\Fig}[1]{{Figure~\ref{fig:#1}}} 
\newcommand{\Tab}[1]{{Table~\ref{tab:#1}}} 
\newcommand{\Thm}[1]{{Theorem~\ref{thm:#1}}} 
\newcommand{\Lem}[1]{{Lemma~\ref{lem:#1}}} 
\newcommand{\Prop}[1]{{Property~\ref{prop:#1}}} 
\newcommand{\Cor}[1]{{Corollary~\ref{cor:#1}}} 
\newcommand{\Def}[1]{{Definition~\ref{def:#1}}} 
\newcommand{\Alg}[1]{{Algorithm~\ref{alg:#1}}} 
\newcommand{\Ex}[1]{{Example~\ref{ex:#1}}} 
\newcommand{\As}[1]{{Assumption~\ref{as:#1}}} 
\newcommand{\Reg}[1]{{R~\ref{reg:#1}}} 
\newcommand{\AlgLine}[2]{{line~\ref{line:#2} of Algorithm~\ref{alg:#1}}}
\newcommand{\AlgLines}[3]{{lines~\ref{line:#2}--\ref{line:#3} of Algorithm~\ref{alg:#1}}}
\fi

\newcommand{\Real}{\mathbb{R}}

\newcommand{\Tra}{^{\sf T}} 

\newcommand{\amp}{\mathop{\:\:\,}\nolimits}

\newcommand{\VI}{\text{VI}}
\newcommand{\One}{\mathbbm{1}}

\newcommand{\V}[1]{{\bm{\mathbf{\MakeLowercase{#1}}}}} 
\newcommand{\VE}[2]{\MakeLowercase{#1}_{#2}} 
\newcommand{\Vhat}[1]{{\bm{\hat \mathbf{\MakeLowercase{#1}}}}} 
\newcommand{\Vtilde}[1]{{\bm{\tilde \mathbf{\MakeLowercase{#1}}}}} 
\newcommand{\Vn}[2]{\V{#1}^{(#2)}} 
\newcommand{\VnE}[3]{{#1}^{(#2)}_{#3}} 

\newcommand{\M}[1]{{\bm{\mathbf{\MakeUppercase{#1}}}}} 
\newcommand{\ME}[2]{\MakeLowercase{#1}_{#2}} 

\newcommand{\Mn}[2]{\M{#1}^{(#2)}} 
\newcommand{\MnE}[3]{\MakeLowercase{#1}^{(#2)}_{#3}} 

\endlocaldefs

\begin{document}

\begin{frontmatter}

\title{Baseline Drift Estimation for Air Quality Data Using Quantile Trend Filtering}
\runtitle{Quantile Trend Filtering}


\author{\fnms{Halley L.} \snm{Brantley}\thanksref{m1}\ead[label=e1]{hlbrantl@ncsu.edu}},
\author{\fnms{Joseph} \snm{Guinness}\thanksref{m2}\ead[label=e2]{guinness@cornell.edu}}
\and
\author{\fnms{Eric C.} \snm{Chi}\corref{}
	\ead[label=e3]{eric$\_$chi@ncsu.edu}\thanksref{m1}}
\affiliation{North Carolina State University\thanksmark{m1} and Cornell University\thanksmark{m2}}
\address{Department of Statistics \\ 2311 Stinson Dr\\ Raleigh, NC 27606\\ \printead{e1}\\
\printead{e3}}

\address{Department of Statistics and Data Science \\ 129 Garden Ave.\\ Ithaca, NY 14853\\ \printead{e2}}

\runauthor{H. L. Brantley et al.}

\begin{abstract}
	We address the problem of estimating smoothly varying baseline trends in time series data. This problem arises in a wide range of fields, including chemistry, macroeconomics, and medicine; however, our study is motivated by the analysis of data from low cost air quality sensors. Our methods extend the quantile trend filtering framework to enable the estimation of multiple quantile trends simultaneously while ensuring that the quantiles do not cross. To handle the computational challenge posed by very long time series, we propose a parallelizable alternating direction method of moments (ADMM) algorithm. The ADMM algorthim enables the estimation of trends in a piecewise manner, both reducing the computation time and extending the limits of the method to larger data sizes. We also address smoothing parameter selection and propose a modified criterion based on the extended Bayesian Information Criterion. Through simulation studies and our motivating application to low cost air quality sensor data, we demonstrate that our model provides better quantile trend estimates than existing methods and improves signal classification of low-cost air quality sensor output.
\end{abstract}



\end{frontmatter}

\section{Introduction}
\label{sec:intro}
\begin{figure}[t!]
	\includegraphics[width = \linewidth]{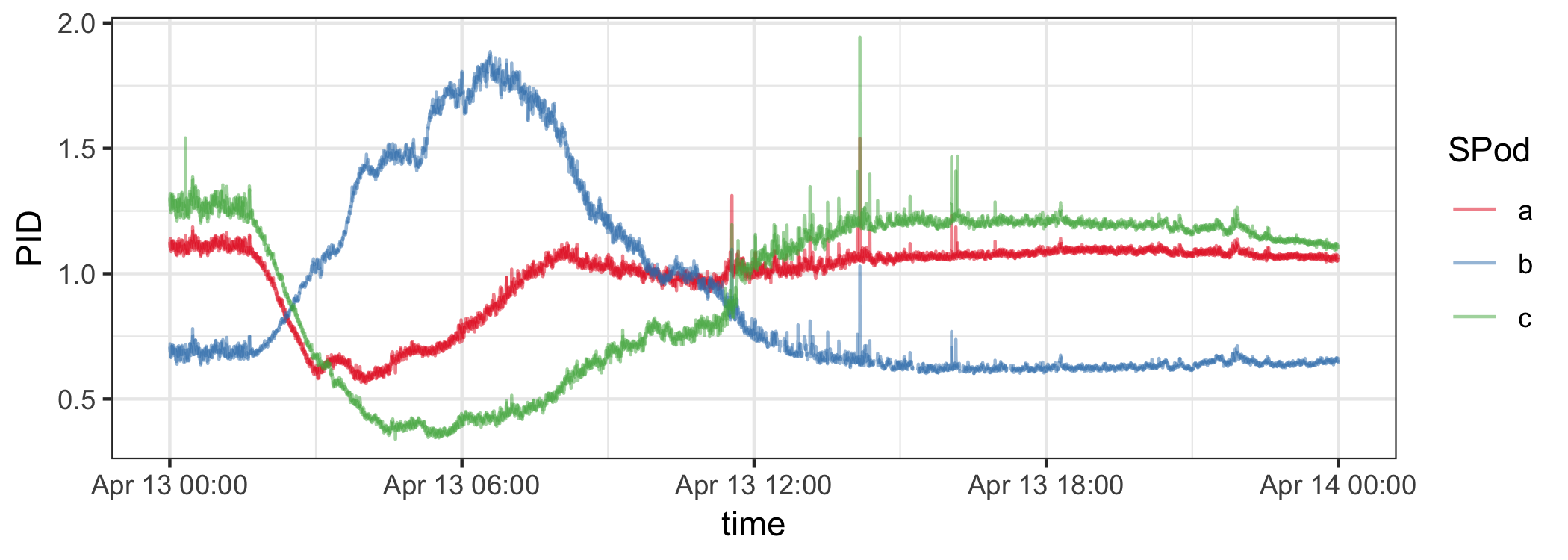}
	\caption{Example of 3 co-located SPod PID sensor readings over a 24 hour period.}
	\label{fig:raw_spod}
\end{figure}

In the last decade, low cost and portable air quality sensors have enjoyed dramatically increased usage. These sensors can provide an un-calibrated measure of a variety of pollutants in near real time, but deriving meaningful information from sensor data remains a challenge \citep{snyder2013changing}. The ``SPod" is a low-cost sensor currently being investigated by researchers at the U.S. Environmental Protection Agency to detect volatile organic compound (VOC) emissions from industrial facilities \citep{thoma2016south}. Due to changes in temperature and relative humidity the output signal exhibits a slowly varying baseline drift on the order of minutes to hours. Figure \ref{fig:raw_spod} provides an example of measurements from three SPod sensors co-located at the border of an industrial facility. All of the sensors respond to the pollutant signal, which is illustrated by the three sharp transient spikes at 11:32, 14:10, and 16:03. However, the baseline drift varies from one sensor to another, obscuring the detection of the peaks that alert the intrusion of pollutants. We show later that by estimating the baseline drift in each sensor and  removing it from the observed signals, peaks can be reliably detected from concordant residual signals from a collection of SPods using a simple data-driven thresholding strategy. Thus, accurately demixing a noisy observed time series into a slowly varying component and a transient component can lead to greatly improved and simplified downstream analysis. 

While this work is motivated by the analysis of data from low cost air quality sensors, the problem of demixing nosy time series into trends and transients is ubiquitous across many fields of study. In a wide range of applications that spans chemistry \citep{Ning2014}, macroeconomics \citep{yamada2017estimating}, environmental science \citep{brantley2014mobile}, and medical sciences \citep{pettersson2013algorithm, marandi2015qualitative}, scalar functions of time $y(t)$ are observed and assumed to be a superposition of an underlying slowly varying baseline trend $\theta(t)$, other more rapidly varying components $s(t)$, and noise. In practice, $y(t)$ is observed at discrete time points $t_1, \ldots, t_n$, and we model the vector of samples $\VE{y}{i} = y(t_i)$ as
\begin{eqnarray*}
	\V{y} & = & \V{\theta} + \V{s} + \V{\varepsilon},
\end{eqnarray*}
where $\VE{\theta}{i} = \theta(t_i)$, $\VE{s}{i} = s(t_i)$, and $\V{\varepsilon} \in \Real^n$ is a vector of uncorrelated noise. For notational simplicity, for the rest of the paper, we assume that the time points take on the values $t_i = i$, but it is straightforward to generalize to an arbitrary grid of time points.

In some applications, the slowly varying component $\V{\theta}$ is the signal of interest, and the transient component $\V{s}$ is a vector of nuisance parameters. In our air quality application, the roles of $\V{\theta}$ and $\V{s}$ are reversed; $\V{s}$ represents the signal of interest and $\V{\theta}$ represents a baseline drift that obscures the identification of the important transient events encoded in $\V{s}$. 

To tackle demixing problems, we introduce a scalable baseline estimation framework by building on \textit{$\ell_1$-trend filtering}, a relatively new nonparametric estimation framework. Our contributions are three-fold.
\begin{itemize}
	\item \cite{Kim2009} proposed using the check function as a possible extension of $\ell_1$-trend filtering but did not investigate it further. Here, we develop the basic $\ell_1$-quantile-trend-filtering framework and extend it to model {\em multiple quantiles simultaneously with non-crossing constraints to ensure validity and improve trend estimates}.
	\item To reduce computation time and extend the method to long time series, we develop a parallelizable ADMM algorithm. The algorithm proceeds by splitting the time domain into overlapping windows, fitting the model separately for each of the windows and reconciling estimates from the overlapping intervals.
	\item Finally, we propose a modified criterion for performing model selection.
\end{itemize}

In the rest of the paper, we detail our quantile trend filtering algorithms (Section \ref{sec:methods}) as well as how to choose the smoothing parameter (Section \ref{sec:lambda_choice}). We demonstrate through simulation studies that our proposed model provides better or comparable estimates of non-parametric quantile trends than existing methods (Section \ref{sec:simluation}). We further show that quantile trend filtering is a more effective method of drift removal for low-cost air quality sensors and results in improved signal classification compared to quantile smoothing splines (Section \ref{sec:application}). Finally, we discuss potential extensions of quantile trend filtering (Section \ref{sec:discussion}).

\section{Baseline Trend Estimation}
\label{sec:methods}

\subsection{Background}

\cite{Kim2009} originally proposed \textit{$\ell_1$-trend filtering} to estimate trends with piecewise polynomial functions, assuming that the observed time series $\V{y}$ consists of a trend $\V{\theta}$ plus uncorrelated noise $\V{\varepsilon}$, namely $\V{y} = \V{\theta} + \V{\varepsilon}$. The estimated trend is the solution to the following convex optimization problem
\begin{eqnarray*}
	\underset{\V{\theta}}{\text{minimize}}\; \frac{1}{2} \lVert \V{y} - \V{\theta} \rVert_2^2 + \lambda \lVert \Mn{D}{k+1}\V{\theta} \rVert_1,
\end{eqnarray*}
where $\lambda$ is a nonnegative regularization parameter, and the matrix $\Mn{D}{k+1} \in \Real^{(n - k -1) \times n}$ is the discrete difference operator of order $k+1$. To understand the purpose of penalizing the 1-norm of $\Mn{D}{k+1}\V{\theta}$ consider the difference operator when $k = 0$.

\begin{eqnarray*}
	\Mn{D}{1} & = & \begin{pmatrix}
		-1 & 1 & 0 & \cdots & 0 & 0 \\
		0 & -1 & 1 & \cdots & 0 & 0 \\
		\vdots & \vdots & \vdots & \ddots & \vdots & \vdots \\
		0 & 0 & 0 & \cdots & -1 & 1 \\
	\end{pmatrix}.
\end{eqnarray*}

Thus, $\lVert \Mn{D}{1}\V{\theta} \rVert_1 = \sum_{i=1}^{n-1} \lvert \theta_i - \theta_{i+1} \rvert$, which is known as the total variation denoising penalty in one dimension in the signal processing literature \citep{Rudin1992} or the fused lasso penalty in the statistics literature \citep{Tibshirani2005}. The penalty term incentivizes solutions which are piecewise constant. For $k \geq 1$, the difference operator $\Mn{D}{k+1} \in \Real^{(n-k-1) \times n}$ is defined recursively as follows
\begin{eqnarray*}
	\Mn{D}{k+1} & = & \Mn{D}{1}\Mn{D}{k}.
\end{eqnarray*}
Penalizing the 1-norm of the vector $\Mn{D}{k+1}\V{\theta}$ produces estimates of $\V{\theta}$ that are piecewise polynomials of order $k$.

\begin{figure}
	\centering
	\subfloat[Slow Trend + Noise]{\label{fig:tpn}
		\includegraphics[width = 0.45\linewidth]{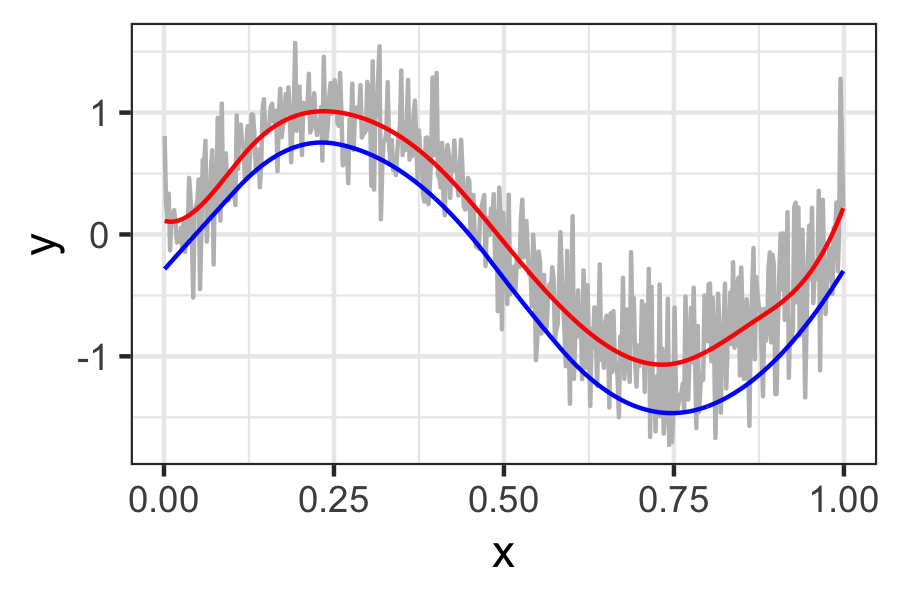}}
	\subfloat[Slow Trend + Spikes + Noise]{\label{fig:tpspn}
		\includegraphics[width = 0.45\linewidth]{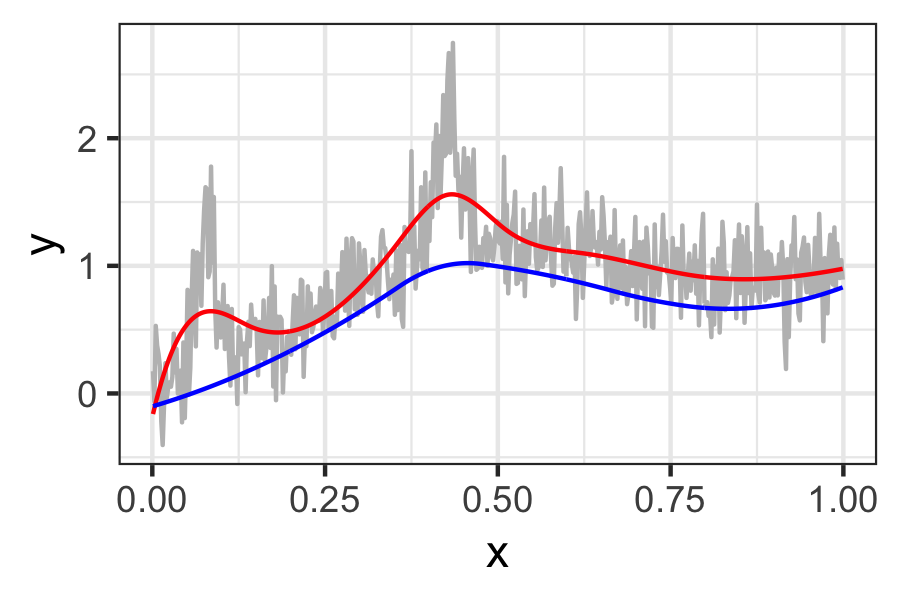}}
	\caption{Examples of trend filtering solutions (red) and 15\textsuperscript{th} quantile trending filtering solution (blue). Standard trend filtering performs well in the no-signal case (a) but struggles to distinguish between the slowly varying trend and the rapidly-varying signal (b). The quantile trend is not affected by the signal and provides an estimate of the baseline.}
\end{figure}

\cite{Tib2014} proved that with a judicious choice of $\lambda$ the trend filtering estimate converges to the true underlying function at the minimax rate for functions whose $k$th derivative is of bounded variation and showed that trend filtering is locally adaptive when the time series consists of only the trend and random noise, which is illustrated in \Fig{tpn}. As noted earlier, in some applications, such as the air quality monitoring problem considered in this paper, the data contain a rapidly varying signal in addition to the slowly varying trend and noise. \Fig{tpspn} shows that standard trend filtering is not designed to distinguish between the slowly varying trend and the rapidly-varying signal, as the smooth component estimate $\V{\theta}$ is biased towards the peaks of the transient components.

To account for the presence of transient components in the observed time series $\V{y}$, we propose quantile trend filtering \Fig{tpspn}. 
To estimate the trend in the $\tau$\textsuperscript{th} quantile, we solve the convex optimization problem
\begin{eqnarray}
\label{eq:quantile_trend}
\underset{\V{\theta}}{\text{minimize}}\; \rho_\tau(\V{y} - \V{\theta}) + \lambda \lVert \Mn{D}{k+1} \V{\theta} \rVert_1,
\end{eqnarray}
where $\rho_\tau(\V{r})$ is the check function
\begin{eqnarray}
\label{eq:check}
\rho_{\tau}(\V{r}) & = & \sum_{i=1}^n \VE{r}{i}(\tau-\One(\VE{r}{i}<0)),
\end{eqnarray}
and $\One(A)$ is 1 if its input $A$ is true and 0 otherwise. Note that we do not explicitly model $\V{s}$. Rather, we focus on estimating $\V{\theta}$. We then estimate $\V{s} + \V{\varepsilon}$ as the difference $\V{y} - \V{\theta}$.

Before elaborating on how we compute our proposed $\ell_1$-quantile trend filtering estimator, we discuss similarities and differences between our proposed estimator and existing quantile trend estimators.

\subsubsection{Relationship to Prior Work}

In this application, as well as those described in \cite{Ning2014}, \cite{marandi2015qualitative}, and \cite{pettersson2013algorithm}, the goal is to estimate the trend in the baseline not the mean. We can define the trend in the baseline as the trend in a low quantile of the data. A variety of methods for estimating quantile trends have already been proposed. \cite{Koenker1978} were the first to propose substituting the sum-of-squares term with the check function  \Eqn{check} to estimate a conditional quantile instead of the conditional mean. Later, \cite{KoenkerNgPortnoy1994} proposed quantile trend filtering with $k = 1$ producing quantile trends that are piecewise linear, but they did not consider extensions to higher order differences. Rather than using the $\ell_1$-norm to penalize the discrete differences, \cite{nychka1995nonparametric} used the smoothing spline penalty based on the square of the $\ell_2$-norm:

\begin{equation*}
\label{eq:smoothingspline}
\sum_{i=1}^n\rho_{\tau}(y(t_i) - \theta(t_i)) + \lambda\int \theta''(t)^2 dt,
\end{equation*}
where $\theta(t)$ is a smooth function of time and $\lambda$ is a tuning parameter that controls the degree of smoothing. \cite{Oh2011} proposed an algorithm for solving the quantile smoothing spline problem by approximating the check function with a differentiable function. \cite{Racine2017} propose a method for estimating quantile trends that does employ the check function. In their method, the response is constrained to follow a location scale model and the conditional quantiles are estimated by combining Gaussian quantile functions with a kernel smoother and solving a local-linear least squares problem.

\subsection{Quantile Trend Filtering}

We combine the ideas of quantile regression and trend filtering. For a single quantile level $\tau$, the quantile trend filtering problem is given in \Eqn{quantile_trend}. As with classic quantile regression, the quantile trend filtering problem is a linear program which can be solved by a number of methods. We want to estimate multiple quantiles simultaneously and to ensure that our quantile estimates are valid by enforcing the constraint that if $\tau_2 > \tau_1$ then $Q(\tau_2) \ge Q(\tau_1)$ where $Q$ is the quantile function of $\V{y}$. Even if a single quantile is ultimately desired, ensuring non-crossing allows information from nearby quantiles to be used to improve the estimates as we will see in the peak detection experiments in \Sec{peak_detection}. Given quantiles $\tau_1 < \tau_2 < \ldots < \tau_J$, the optimization problem becomes
\begin{eqnarray}
	\label{eq:noncross_trend}
	&&\underset{\M{\Theta} \in \mathcal{C}}{\text{minimize}}\; \sum_{j=1}^J \left [\rho_{\tau_j}(\V{y} - \V{\theta}_{j}) +
	\lambda_j \lVert \Mn{D}{k+1} \V{\theta}_j \rVert_1 \right ] 
\end{eqnarray}
where $\M{\Theta} = \begin{pmatrix} \V{\theta}_1 & \V{\theta}_2 & \cdots & \V{\theta}_J \end{pmatrix} \in \Real^{n \times J}$ is a matrix whose $j$th column corresponds to the $j$th quantile signal $\V{\theta}_j$ and the set $\mathcal{C} = \{\M{\Theta} \in \Real^{n \times J}: \ME{\theta}{ij} \leq \ME{\theta}{ij'} \;\text{for $j \leq j'$}\}$ encodes the non-crossing quantile constraints. The additional non-crossing constraints are linear inequalities involving the parameters, so the non-crossing quantile trends can still be estimated by a number of available linear programming solvers. We allow for the possibility that the degree of smoothness in the trends varies by quantile by allowing the smoothing parameter to vary with quantile as well. In the rest of this paper, we use $k=2$ to produce piecewise quadratic polynomials and report numerical results using the commercial solver Gurobi \citep{gurobi} and its R package implementation. However, we could easily substitute a free solver such as the Rglpk package by \cite{rglpk}.

\subsection{ADMM for Big Data}

 As the size of the data increases, computation time becomes prohibitive. In our application to air quality sensor data, measurements are recorded every second resulting in 86,400 observations per day. This number of observations is already too large to use with currently available R packages used for estimating quantile trends \citep{fields, quantreg}. To our knowledge, no one has addressed the problem of finding smooth quantile trends of series that are too large to be processed simultaneously. We propose a divide-and-conquer approach via an ADMM algorithm for solving large problems in a piecewise fashion. 
 
 \begin{figure}[!t]
 	\centering
 	\includegraphics[width = 0.8\linewidth]{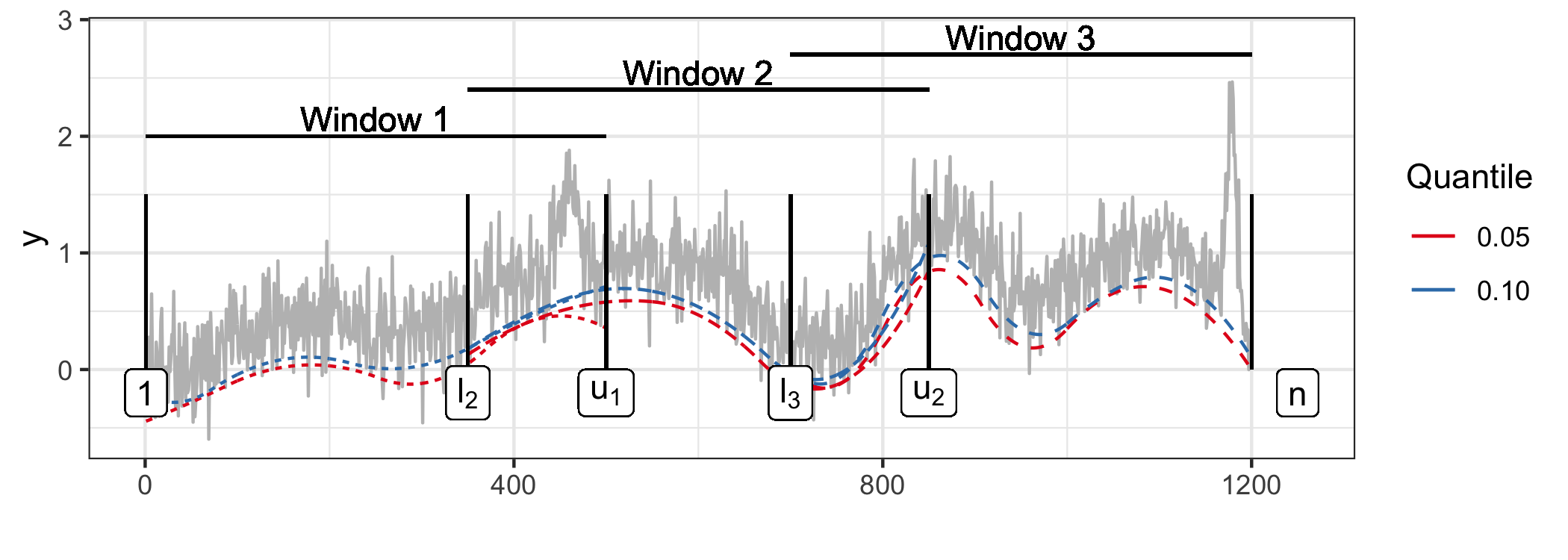}
 	\caption{Window boundaries and trends fit {\em separately} in each window. Each window's trend estimate is plotted in a different line type.}
 	\label{fig:windows}
 \end{figure}
 
\subsubsection{Formulation}
 
 To decrease computation time and extend our method to larger problems, we divide our observed series $\VE{y}{i}$ with $i = 1, \ldots, n$ into $W$ overlapping windows of observations, defining the vector of sequential elements indexed from $l_w$ to $u_w$ as $\Vn{y}{w} = \{y_{l_w}, y_{l_w + 1}, \ldots, y_{u_w -1}, y_{u_w}\}$, with 
 \begin{eqnarray*}
 	1 = l_{1} < l_{2} < u_{1} < l_{3} < u_{2} < l_{4} < u_{3} < \cdots < u_{W} = n.
 \end{eqnarray*}
 We define $n_w = u_w-l_w+1$ so that $\Vn{y}{w} \in \Real^{n_w}$. \Fig{windows} shows an example of 1200 observations being mapped into three equally sized overlapping windows of observations. While the overlapping trend estimates between $l_2$ and $u_2$ do not vary dramatically, the difference is more pronounced in the trend in the 5\textsuperscript{th} quantile between $l_3$ and $u_2$. Thus, we need a way of enforcing estimates to be identical in the overlapping regions. 
 
 Given quantiles $\tau_1 < \cdots < \tau_J$, we introduce dummy variables $\Vn{\theta}{w}_{j} \in \Real^{n_w}$ as the value of the $\tau_j$\textsuperscript{th} quantile trend in window $w$. We then ``stitch" together the $W$ quantile trend estimates into consensus over the overlapping regions by introducing the constraint $\VnE{\theta}{w}{ij} = \VE{\theta}{i + l_w - 1,j}$ for $i = 1, \ldots, n_w$ and for all $j$. Let $\Mn{\Theta}{w}$ be the matrix whose $j$th column is $\Vn{\theta}{w}_{j}$. Then we can write these constraints more concisely as $\Mn{\Theta}{w} = \Mn{U}{w}\M{\Theta}$, where $\Mn{U}{w} \in \{0, 1\}^{n_w \times n}$ is a matrix that selects rows of $\M{\Theta}$ corresponding to the $w$th window, namely
 \begin{eqnarray*}
 	\Mn{U}{w} & = & \begin{pmatrix}
 		\V{e}_{l_w}\Tra \\
 		\vdots \\
 		\V{e}_{u_w}\Tra
 	\end{pmatrix},
 \end{eqnarray*}
 where $\V{e}_i \in \Real^{n}$ denotes the $i$th standard basis vector. Furthermore, let $\iota_{\mathcal{C}}$ denote the indicator function of the non-crossing quantile constraint, namely $\iota_{\mathcal{C}}(\M{\Theta})$ is zero if $\M{\Theta} \in \mathcal{C}$ and infinity otherwise. Our windowed quantile trend optimization problem can then be written as
 \begin{equation}
 \label{eq:quantile_windows}
 \begin{split}
 &\text{minimize}\; \sum_{w=1}^W \left \{\sum_{j=1}^J \left [\rho_{\tau_j}(\Vn{y}{w} - \Vn{\theta}{w}_{j}) +
 \lambda_j \lVert \Mn{D}{k+1} \Vn{\theta}{w}_{j} \rVert_1 \right ] + \iota_\mathcal{C}(\Mn{\Theta}{w}) \right\}\\
 &\text{subject to} \qquad \Mn{\Theta}{w} = \Mn{U}{w}\M{\Theta} \;\; \text{ for } w = 1, \ldots, W.
 \end{split}
 \end{equation}
  
\begin{figure}
 	\centering
 	\includegraphics[width = 0.8\linewidth]{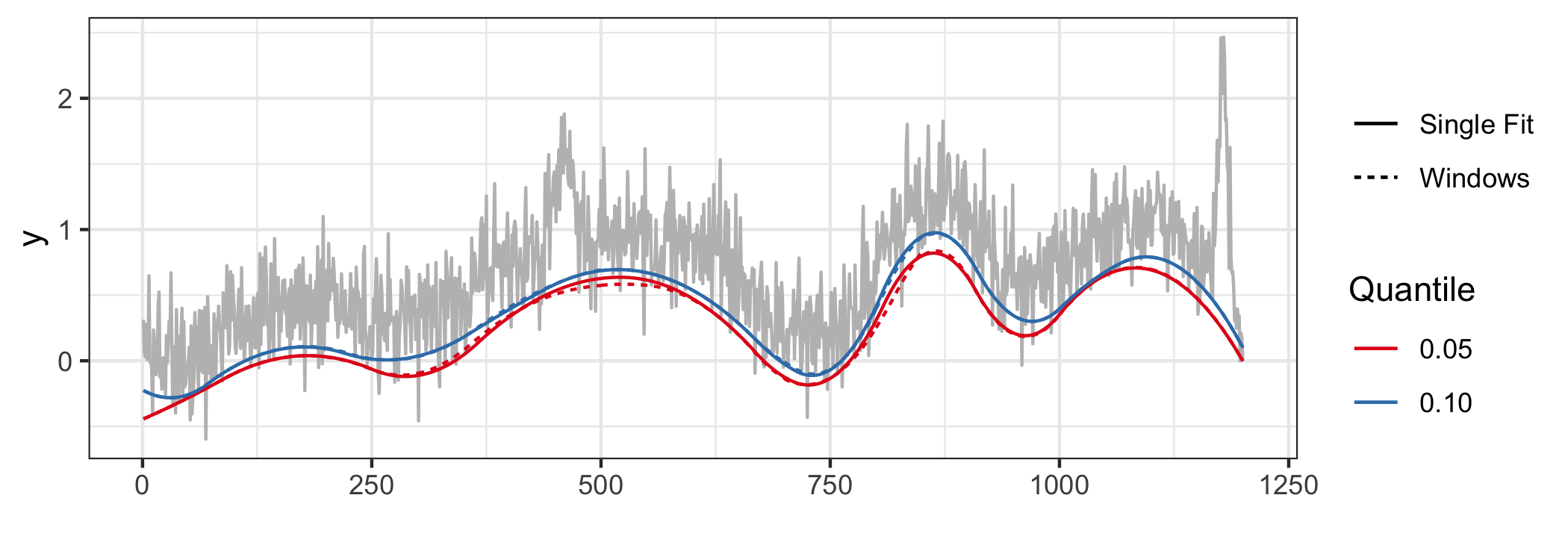}
 	\caption{Trend fit with our ADMM algorithm with 3 windows (converged in 7 iterations), compared to trend from simultaneous fit.} 
 	\label{fig:windowsfit}
 \end{figure}

 The solution to \Eqn{quantile_windows} is {\em not} identical to the solution to \Eqn{noncross_trend} because of double counting of the overlapping sections. The solutions are very close, however, and the differences are essentially immaterial concerning downstream analysis. \Fig{windowsfit} provides an illustration of the trends estimated using multiple windows compared with the trends estimated using a single window; estimates using multiple and single windows are nearly indistinguishable.

\subsubsection{Algorithm}
The ADMM algorithm \citep{gabay1975dual, glowinski1975approximation} is  described in greater detail by \cite{boyd2011distributed}, but we briefly review how it can be used to iteratively solve the following equality constrained optimization problem which is a more general form of \Eqn{quantile_windows}.

\begin{equation}
\label{eq:split_objective}
\begin{split}
&\text{minimize} \; f(\V{\phi}) + g(\Vtilde{\phi}) \\ 
&\text{subject to} \; \M{A}\V{\phi} + \M{B}\Vtilde{\phi} = \V{c}.
\end{split}
\end{equation}
Recall that finding the minimizer to an equality constrained optimization problem is equivalent to the identifying the saddle point of the Lagrangian function associated with the problem \Eqn{split_objective}. ADMM seeks the saddle point of a related function called the augmented Lagrangian,
\begin{eqnarray*}
	\mathcal{L}_{\gamma}(\V{\phi},\Vtilde{\phi},\V{\omega}) & = & f(\V{\phi}) + g(\Vtilde{\phi}) + \langle \V{\omega}, \V{c} - \M{A}\V{\phi} - \M{B}\Vtilde{\phi} \rangle
	+ \frac{\gamma}{2} \lVert \V{c} - \M{A}\V{\phi} - \M{B}\Vtilde{\phi} \rVert_2^2,
\end{eqnarray*}
where the dual variable $\V{\omega}$ is a vector of Lagrange multipliers and $\gamma$ is a nonnegative tuning parameter. When $\gamma = 0$, the augmented Lagrangian coincides with the ordinary Lagrangian.

ADMM minimizes the augmented Lagrangian one block of variables at a time before updating the dual variable $\V{\omega}$. This yields the following sequence of updates at the $(m+1)$\textsuperscript{th} ADMM iteration
\begin{equation}
\label{eq:admm_updates}
\begin{split}
\V{\phi}_{m+1} & \amp = \amp \underset{\V{\phi}}{\arg\min}\; \mathcal{L}_\gamma(\V{\phi},\Vtilde{\phi}_{m}, \V{\omega}_{m}) \\
\Vtilde{\phi}_{m+1} & \amp = \amp \underset{\Vtilde{\phi}}{\arg\min}\; \mathcal{L}_\gamma(\V{\phi}_{m+1},\Vtilde{\phi}, \V{\omega}_{m}) \\
\V{\omega}_{m+1} & \amp = \amp \V{\omega}_{m} + \gamma(\V{c} - \M{A}\V{\phi}_{m+1} - \M{B}\Vtilde{\phi}_{m+1}).
\end{split}
\end{equation}

Returning to our constrained windows problem giving in \Eqn{quantile_windows}, let $\Mn{\Omega}{w}$ denote the Lagrange multiplier matrix for the $w$th consensus constraint, namely $\Mn{\Theta}{w} = \Mn{U}{w}\M{\Theta}$, and let $\Vn{\omega}{w}_{j}$ denote its $j$th column.

The augmented Lagrangian is given by
\begin{eqnarray*}
	\mathcal{L}(\M{\Theta}, \{\Mn{\Theta}{w}\}, \{\Mn{\Omega}{w}\}) & = & \sum_{w=1}^W \mathcal{L}_w(\M{\Theta}, \Mn{\Theta}{w}, \Mn{\Omega}{w}),
\end{eqnarray*}
where
\begin{eqnarray*}
	\mathcal{L}_w(\M{\Theta}, \Mn{\Theta}{w}, \Mn{\Omega}{w}) & = & \sum_{j=1}^J \biggl [\rho_{\tau_j}(\Vn{y}{w} - \Vn{\theta}{w}_j)+\lambda_j \lVert \Mn{D}{k+1}\Vn{\theta}{w}_j\rVert_1 \\
	&& + \; (\Vn{\theta}{w}_j - \Mn{U}{w}\V{\theta}_{j})\Tra\Vn{\omega}{w}_{j} +
	\frac{\gamma}{2}\lVert \Vn{\theta}{w}_j - \Mn{U}{w}\V{\theta}_{j}\rVert_2^2 \biggr ] + \iota_\mathcal{C}(\Mn{\Theta}{w}),
\end{eqnarray*}
where $\gamma$ is a positive tuning parameter.

The ADMM algorithm alternates between updating the consensus variable $\M{\Theta}$, the window variables $\{\Mn{\Theta}{w}\}$, and the Lagrange multipliers $\{\Mn{\Omega}{w}\}$.
At the $(m+1)$\textsuperscript{th} iteration, we perform the following sequence of updates

\begin{eqnarray*}
	\M{\Theta}_{m+1} & = & \underset{\M{\Theta}}{\arg\min}\; \mathcal{L}(\M{\Theta}, \{\Mn{\Theta}{w}_m\}, \{\Mn{\Omega}{w}_m\}) \\
	\Mn{\Theta}{w}_{m+1} & = & \underset{\{\Mn{\Theta}{w}\}}{\arg\min}\; \mathcal{L}(\M{\Theta}_{m+1}, \{\Mn{\Theta}{w}\}, \{\Mn{\Omega}{w}_m\}) \\
\end{eqnarray*}

{\bf Updating $\M{\Theta}$: } Some algebra shows that,  defining $i_w = i-l_w+1$, updating the consensus variable step is computed as follows.

\begin{align}
\label{eq:update_consensus}
~~~~\ME{\theta}{ij}	 &=& \begin{cases}
\frac{1}{2}\left (\MnE{\theta}{w-1}{i_{w-1}j} + \MnE{\theta}{w}{i_wj} \right)
-
\frac{1}{2\gamma} \left (\MnE{\omega}{w-1}{i_{w-1}j} + \MnE{\omega}{w}{i_{w}j} \right)
& \mbox{if~} l_{w} \le i \le u_{w-1},  \\
\MnE{\theta}{w}{i_{w}j} & \mbox{if~} u_{w-1} < i \le l_{w+1}  \\
\frac{1}{2} \left ( \MnE{\theta}{w}{i_{w}j} + \MnE{\theta}{w+1}{i_{w+1}j} \right )
-
\frac{1}{2\gamma}	\left (\MnE{\omega}{w}{i_{w}j} + \MnE{\omega}{w+1}{i_{w+1}j} \right )
& \mbox{if~} l_{w+1} < i \le u_{w}
\end{cases},
\end{align}

The consensus update \Eqn{update_consensus} is rather intuitive. We essentially average the trend estimates in overlapping sections of the windows, subject to some adjustment by the Lagrange multipliers, and leave the trend estimates in non-overlapping sections of the windows untouched.
For notational ease, we write the consensus update \Eqn{update_consensus} compactly as $\M{\Theta} = \psi(\{\Mn{\Theta}{w}\},\{\Mn{\Omega}{w}\})$.

{\bf Updating $\{\Mn{\Theta}{w}\}$: } We then estimate the trend separately in each window, which can be done in parallel, while penalizing the differences in the overlapping pieces of the trends  as outlined in \Alg{admm}. The use of the Augmented Lagrangian converts the problem of solving a potentially large linear program into a solving a collection of smaller quadratic programs. The \texttt{gurobi} R package \citep{gurobi} can solve quadratic programs in addition to linear programs, but we can also use the free R package \texttt{quadprog} \citep{quadprog}.

\begin{algorithm}[t]
	\caption{ADMM algorithm for quantile trend filtering with windows}\label{alg:admm}
	\begin{algorithmic}
		\State Define $\M{D} = \Mn{D}{k+1}$.
		\State \textbf{initialize:}
		\For{$w = 1, \ldots, W$}
		\State $\Mn{\Theta}{w}_0 \gets \underset{\Mn{\Theta}{w} \in \mathcal{C}}{\arg \min}\;
		\sum_{j=1}^J\rho_{\tau_j}(\Vn{y}{w} - \Vn{\theta}{w}_{j})+\lambda \lVert \M{D}\Vn{\theta}{w}_{j}\rVert_1$
		\State $\Mn{\Omega}{w}_0 \gets \M{0}$
		\EndFor
		\State $m \gets 0$
		\Repeat{}
		\State $\M{\Theta}_{m+1} \gets \psi(\{\Mn{\Theta}{w}_m\}, \{\Mn{\Omega}{w}_m\})$
		\For{$w = 1, \ldots, W$}
		\State $\Mn{\Theta}{w}_{m+1} \gets \underset{\Mn{\Theta}{w}}{\arg\min}\; \mathcal{L}_w(\M{\Theta}_{m+1}, \Mn{\Theta}{w}, \Mn{\Omega}{w}_{m})$
		\State
		$\Mn{\Omega}{w}_{m+1} \gets \Mn{\Omega}{w}_m + \gamma(\Mn{\Theta}{w}_{m+1} - \Mn{U}{w}\M{\Theta}_{m+1})$
		\EndFor
		\State $m \gets m + 1$
		\Until {convergence}
		\State \textbf{return} $\M{\Theta}_m$
	\end{algorithmic}
\end{algorithm}

\Alg{admm} has the following convergence guarantees.
\begin{proposition}
	\label{prop:convergence}
	Let $\{\{\Mn{\Theta}{w}_m\}, \M{\Theta}_m\}$ denote the $m$th collection of iterates generated by \Alg{admm}. Then (i)
	$\lVert \Mn{\Theta}{w}_m - \Mn{U}{w}\M{\Theta}_m \rVert_{\text{F}} \rightarrow 0$ and (ii) $p_m \rightarrow p^\star$, where $p^\star$ is the optimal objective function value of
	\Eqn{quantile_windows} and $p_m$ is the objective function value of \Eqn{quantile_windows} evaluated at $\{\{\Mn{\Theta}{w}_m\}, \M{\Theta}_m\}$.
\end{proposition}
The proof of \Prop{convergence} is a straightforward application of the convergence result presented in Section 3.2 of \cite{boyd2011distributed}.

To terminate our algorithm, we use the stopping criteria described by \cite{boyd2011distributed}. The criteria are based on the primal and dual residuals, which represent the residuals for primal and dual feasibility, respectively. The primal residual at the $m$th iteration,

\begin{eqnarray*}
	r_{\text{primal}}^{m} & = & \sqrt{\sum_{w=1}^W\lVert\Mn{\Theta}{w}_{m} - \Mn{U}{w}\M{\Theta}_m\rVert_{\text{F}}^2},
\end{eqnarray*}
represents the difference between the trend values in the windows and the consensus trend value. The dual residual at the $m$th iteration,

\begin{eqnarray*}
	r_{\text{dual}}^{m} & = & \gamma\sqrt{\sum_{w=1}^W \lVert\M{\Theta}_m - \M{\Theta}_{m-1}\rVert_{\text{F}}^2},
\end{eqnarray*}
represents the change in the consensus variable from one iterate to the next. The algorithm is stopped when

\begin{eqnarray}
	r_{\text{primal}}^{m} & < &\epsilon_{\text{abs}}\sqrt{nJ} + \epsilon_{\text{rel}}\,\underset{w}{\max} \left[\max
	\left\{\lVert\Mn{\Theta}{w}_m \rVert_{\text{F}}, \lVert \M{\Theta}_{m} \rVert_{\text{F}} \right\}\right] \nonumber \\
	r_{\text{dual}}^{m} & < & \epsilon_{\text{abs}}\sqrt{nJ} + \epsilon_{\text{rel}}\,\sqrt{\sum_{w=1}^W
		\lVert \Mn{\Omega}{w}_m \rVert_{\text{F}}^2}.
\label{eq:stopping}
\end{eqnarray}

\begin{figure}[!t]
	\centering
	\includegraphics[width = 0.7\linewidth]{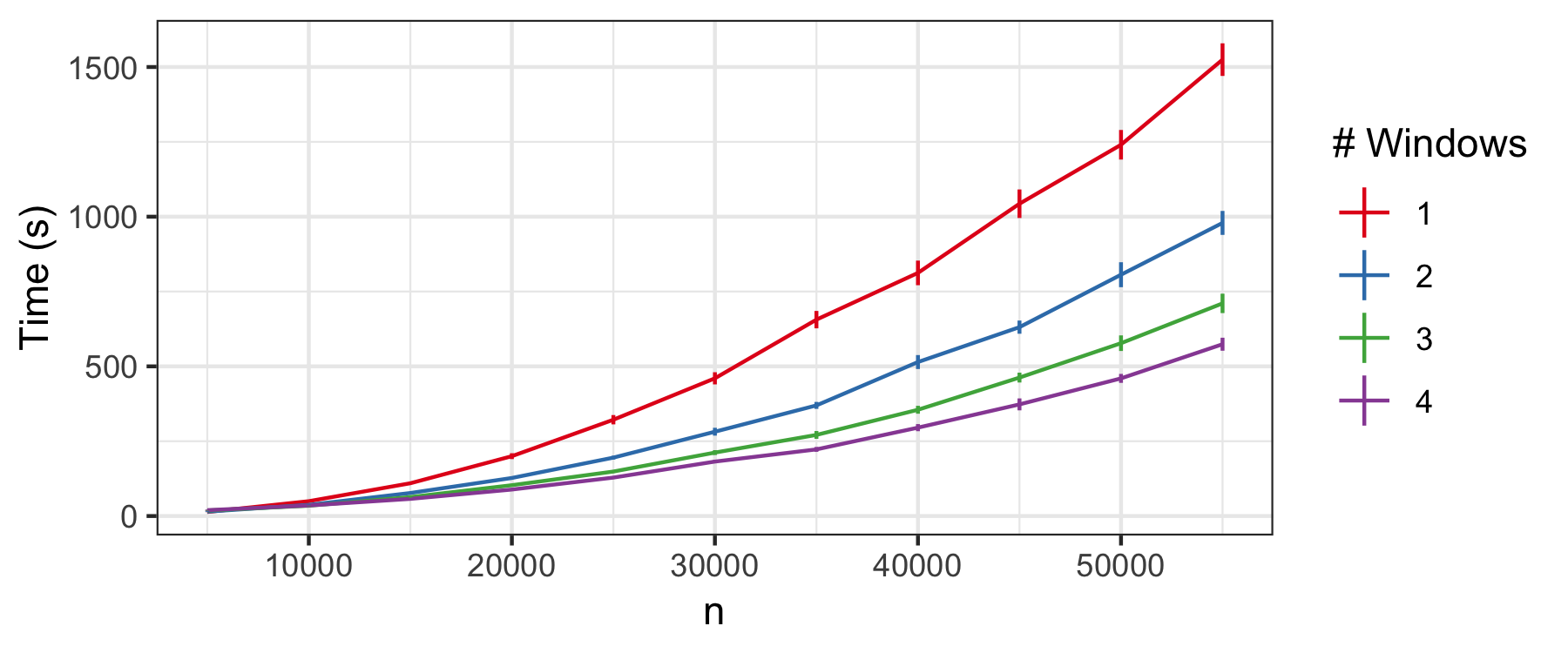}
	\caption{Timing experiments comparing quantile trend filtering with varying numbers of windows by data size.}
	\label{fig:timing}
\end{figure}

The quantile trend filtering problem for a single window is a linear program with $N\times J$ parameters (number of observations by number of quantiles), which can be solved in computational time proportional to $(NJ)^3$.  Consequently, solving a large problem using \Alg{admm} should require less computational time than solving \Eqn{noncross_trend}, even if the sub-problems are solved sequentially. We demonstrate the advantages of \Alg{admm} through timing experiments (\Fig{timing}). For each data size, 25 datasets were simulated using the peaks simulation design described below. Trends for the fifth, tenth, and fifteenth quantiles were fit simultaneously using $\lambda_j = n/5$ for all $j$. We use from one to four windows for each data size with an overlap of 500. \Alg{admm} was the stopped when \Eqn{stopping} was satisfied, defining $\epsilon_{abs} = 0.01$ and $\epsilon_{rel} = 0.001$. \Fig{timing} shows that using 4 windows instead of one on data sizes of 55,000 provides a factor of 3 decrease in computation time. The timing experiments were conducted on an Intel Xeon based Linux cluster using two processor cores.

\section{Model Selection}
\label{sec:lambda_choice}

An important practical issue in baseline estimation is the choice of the regularization parameter $\lambda$, which controls the degree of smoothness in $\V{\theta}$. In this section, we introduce four methods for choosing $\lambda$. The first is a validation based approach; the latter three are based on information criteria. Each of the criteria we compare is calculated for a single quantile ($\tau_j$). Rather than combine results over quantiles, we allow the value of $\lambda$ to vary by quantile resulting in $\V{\lambda} = \{\lambda_1, ..., \lambda_J\}$. To choose the best value for each $\lambda_j$, we first estimate all of the quantile trends using $\lambda_j = \lambda$ for all $j$ over a grid of values for $\lambda$. We then determine the $\lambda_j$ that maximizes the criteria chosen evaluated using $\tau_j$. Finally, we re-estimate the non-crossing trends with the optimal values for $\lambda_j$. A more thorough approach would involve fitting the model on a J dimensional grid of values for $\V{\lambda}$ but this is computationally infeasible.      

\subsection{Validation}
Our method can easily handle missing data by defining the check loss function to output 0 for missing values. 
Specifically, we use the following modified function $\tilde{\rho}_\tau$ in place of the $\rho_\tau$ function given in \Eqn{check}
\begin{eqnarray}
\label{eq:modcheck}
\tilde{\rho}_{\tau}(\V{r}) & = & \sum_{i \not\in \mathcal{V}} \VE{r}{i}(\tau-\One(\VE{r}{i}<0)),
\end{eqnarray}
where $\mathcal{V}$ is a held-out validation subset of $\{1, \ldots, n\}$ and solve the problem

\begin{eqnarray}
\label{eq:validate}
&&\underset{\M{\Theta} \in \mathcal{C}}{\text{minimize}}\; \sum_{j=1}^J \left [\tilde{\rho}_{\tau_j}(\V{y} - \V{\theta}_{j}) +
\lambda_j \lVert \Mn{D}{k+1} \V{\theta}_j \rVert_1 \right ],
\end{eqnarray}
which can be solved via \Alg{admm} with trivial modification to the quadratic program sub-problems. For each quantile level $j$, we select the $\lambda_j$ that minimizes the hold-out prediction error quantified by $\breve{\rho}_{\tau_j}(\V{y} - \Vhat{\theta}_j(\lambda_j))$ where $\Vhat{\theta}_j(\lambda_j)$ is the solution to \Eqn{validate} and
\begin{eqnarray*}
	\breve{\rho}_{\tau}(\V{r}) & = & \sum_{i \in \mathcal{V}} \VE{r}{i}(\tau-\One(\VE{r}{i}<0)).
\end{eqnarray*}

\subsection{Information Criteria}
\cite{KoenkerNgPortnoy1994} addressed the choice of regularization parameter by proposing the Schwarz criterion for the selection of $\lambda$
\begin{eqnarray*}
	\label{eq:SIC}
	\mbox{SIC}(p_{\lambda}, \tau_j) & = & \log\left[\frac{1}{n}\rho_{\tau_j}(\V{y} - \V{\theta}_j)\right] + \frac{1}{2n}p_{\lambda}\log n.
\end{eqnarray*}
where $p_{\lambda} = \sum_i \One(\VE{y}{i} = \widehat{\theta}_i)$ is the number of non-interpolated points, which can be thought of as active knots. Equivalently, $p_{\lambda}$ can be substituted with the number of non-zero components in $\M{D}^{(k+1)}\Vhat{\theta}_j$ which we denote $\nu$ and have found to be more numerically stable.  The SIC is based on the traditional Bayesian Information Criterion (BIC) which is given by
\begin{equation}
\mbox{BIC}(\nu) = -2\log(L\{\Vhat{\theta}\}) + \nu\log n
\end{equation}
where $L$ is the likelihood function. If we take the approach used in Bayesian quantile regression \citep{yu2001bayesian}, and view minimizing the check function as maximizing the asymmetric Laplace likelihood,

\begin{eqnarray*}
	L(\V{y} \mid \V{\theta}) & = & \left[\frac{\tau^n(1-\tau)}{\sigma}\right]^n\exp\left\{-\rho_\tau\left(\frac{\V{y} - \V{\theta}}{\sigma}\right)\right\},
\end{eqnarray*}
we can compute the BIC as
\begin{eqnarray*}
	\mbox{BIC}(\nu, \tau_j) = \frac{2}{\sigma}\rho_{\tau_j}(\V{y}-\Vhat{\theta}_j) + \nu\log n
\end{eqnarray*}
where $\Vhat{\theta}$ is the estimated trend, and $\nu$ is the number of non-zero elements of $\Mn{D}{k+1}\Vhat{\theta}$. We can choose any $\sigma>0$ and have found empirically that $\sigma =  \frac{1-|1-2\tau|}{2}$ produces stable estimates.

\cite{chen2008} proposed the extended Bayesian Information Criteria (eBIC), specifically designed for large parameter spaces.
\begin{eqnarray*}
	\label{eq:eBIC}
	\mbox{eBIC}_{\gamma}(\nu) & = & -2\log(L\{\Vhat{\theta}\}) + \nu\log n  + 2\gamma\log{P \choose \nu},~~\gamma \in [0,1]
\end{eqnarray*}
where $P$ is the total number of possible parameters and $\nu$ is the number of non-zero parameters included in given mod. We used this criteria with $\gamma = 1$, and $P=n-k-1$. In the simulation study, we compare the performance of the SIC, scaled eBIC (with $\sigma$ defined above), and validation methods. 

\section{Simulation Studies}
\label{sec:simluation}

We conduct two simulation studies to compare the performance of our quantile trend filtering method and regularization parameter selection criteria with previously published methods. The first study compares the method's ability to estimate quantiles when the observed series consists of a smooth trend plus independent error, but does not contain transient components. The second study is based on our application and compares the method's ability to estimate baseline trends and enable peak detection when the time series contains a non-negative, transient signal in addition to the trend and random component.

We compare three criteria for choosing the smoothing parameter for quantile trend filtering:  $\V{\lambda}$ chosen using SIC \Eqn{SIC} (\texttt{detrendr\_SIC}); $\V{\lambda}$ chosen using the validation method with the validation set consisting of every 5th observation (\texttt{detrendr\_valid}); and $\V{\lambda}$ chosen using the proposed eBIC criterion \Eqn{eBIC} (\texttt{detrendr\_eBIC}). For the second study we also examine the effect of the non-crossing quantile constraint by estimating the quantile trends separately and choosing $\V{\lambda}$ using eBIC (\texttt{detrendr\_Xing}). We do not include \texttt{detrendr\_Xing} in the first study because the difference in quantiles is large enough that we would not expect the non-crossing constraint to make a difference. 

We also compare the performance of our quantile trend filtering method with three previously published methods, none of which guarantee non-crossing quantiles: 

\begin{itemize}
	\item \texttt{npqw}: The local linear quantile method (quantile-ll) described in \cite{Racine2017}. Code was obtained from the author.
	\item \texttt{qsreg}: Quantile smoothing splines described in \cite{Oh2004period} and \cite{nychka1995nonparametric} and implemented in the \texttt{fields} R package \citep{fields}. The regularization parameter was chosen using generalized cross-validation.
	\item \texttt{rqss}: Quantile trend filtering with $k=1$ available in the \texttt{quantreg} package and described in \cite{KoenkerNgPortnoy1994}. The regularization parameter is chosen using a grid search and minimizing the SIC \Eqn{SIC} as described in \cite{KoenkerNgPortnoy1994}.
\end{itemize}

\subsection{Estimating Quantiles}
\begin{figure}
	\includegraphics[width=.27\linewidth]{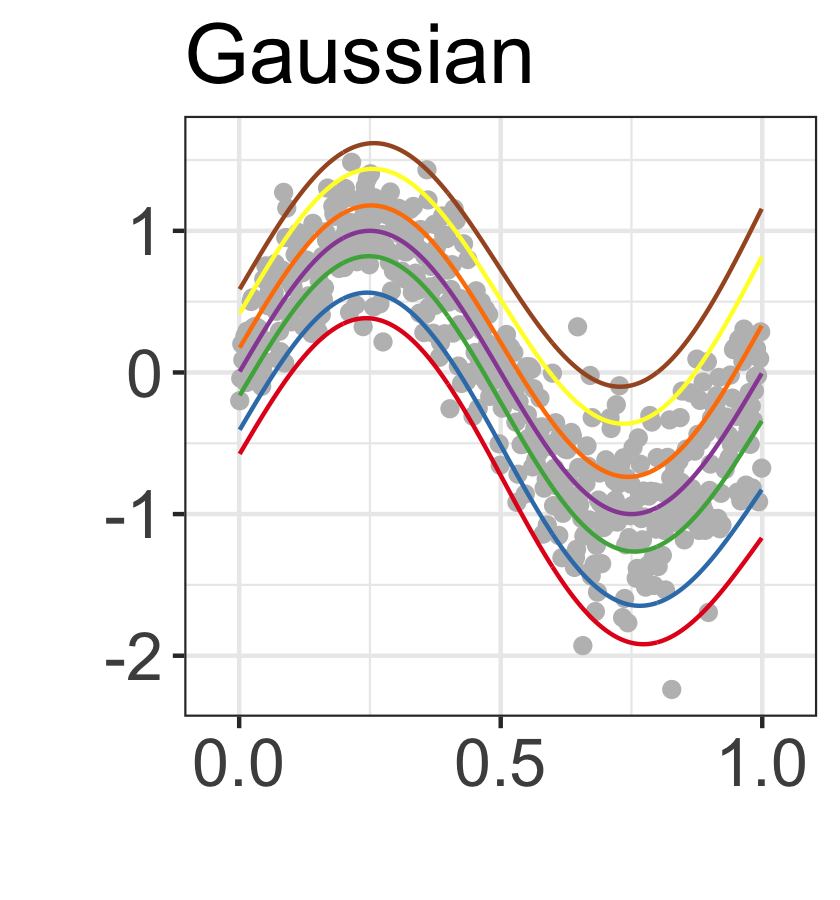}
	\includegraphics[width=.29\linewidth]{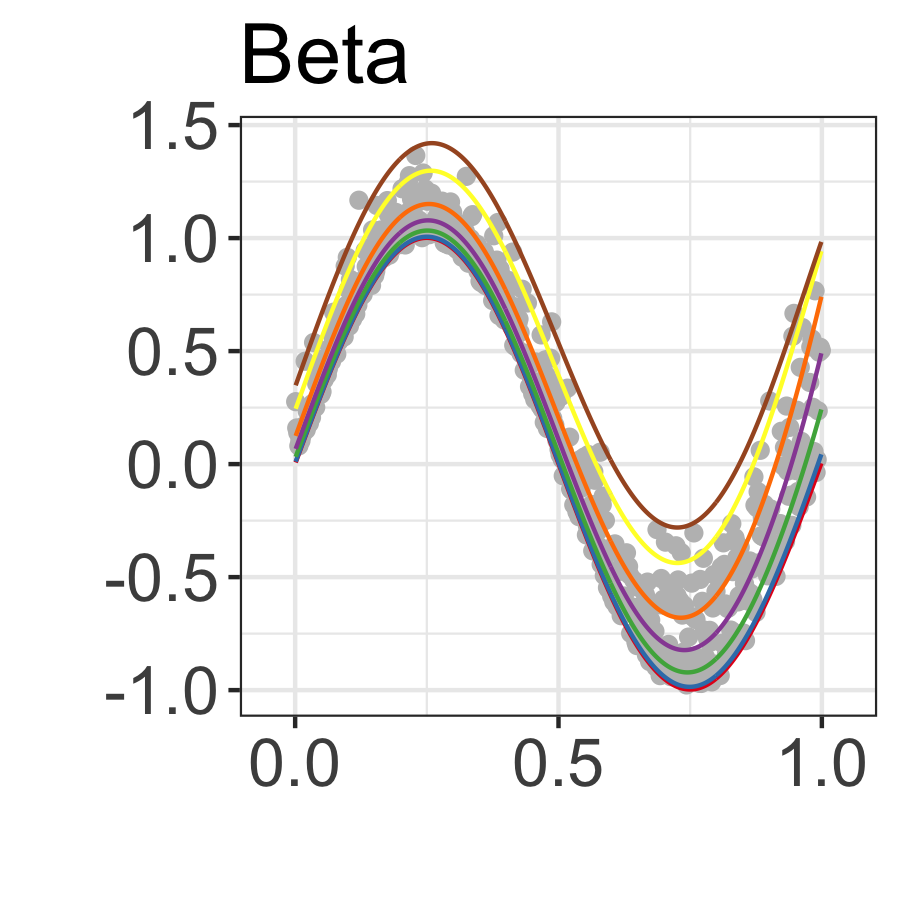}
	\includegraphics[width=.44\linewidth]{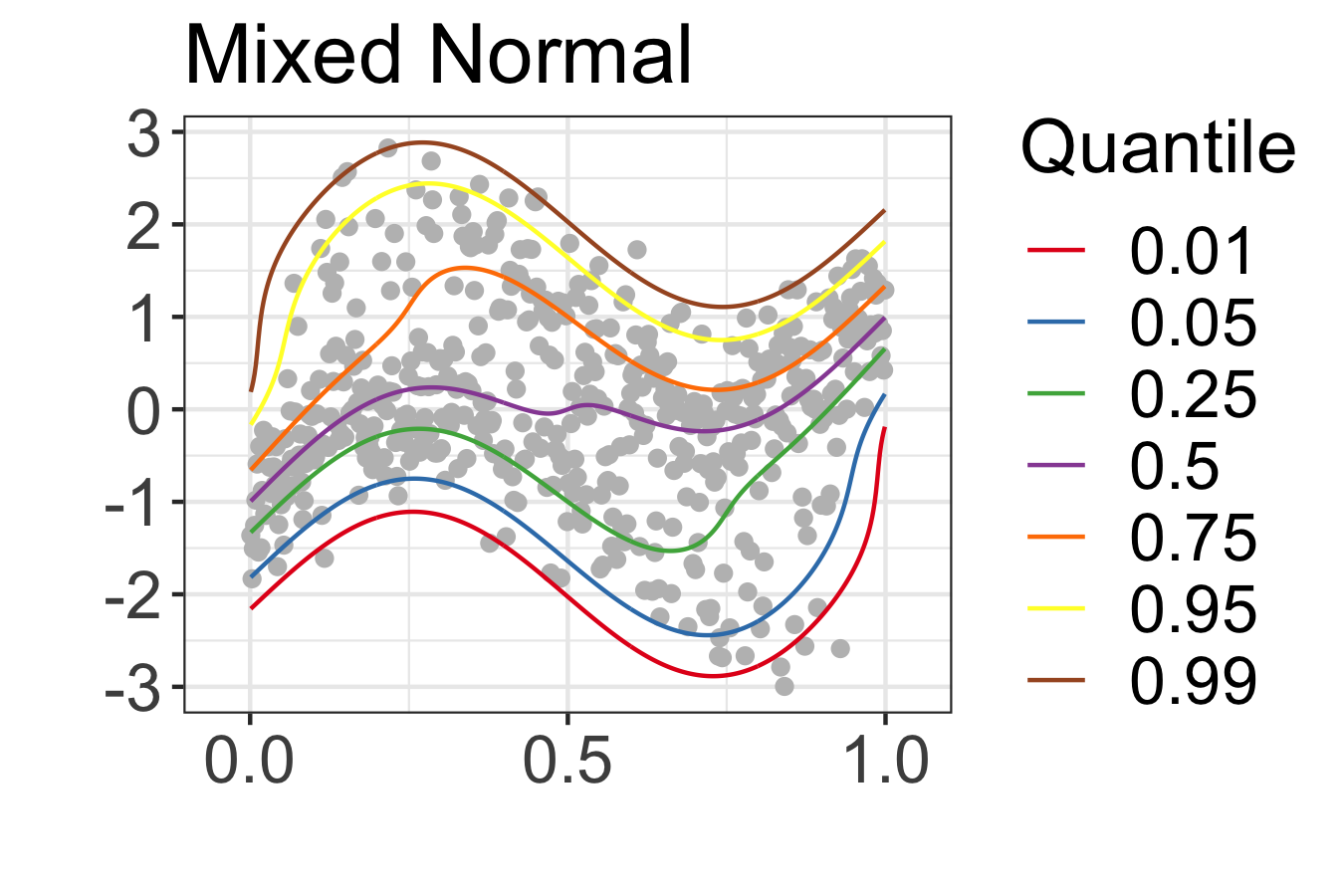}
	\caption{Simulated data with true quantile trends.}
	\label{fig:sim1}
\end{figure}

To compare performance in estimating quantile trends in the absence of a signal component, three simulation designs from \cite{Racine2017} were considered. For all designs $t = 1, \ldots, n$,  $x(t) = t/n$, and the response $\V{y}$ was generated as
\begin{eqnarray*}
	y(t) & = & \sin(2\pi x(t)) + \epsilon(x(t))
\end{eqnarray*}
The errors were simulated as independent draws from the following distributions:
\begin{itemize}
	\item Gaussian: $\epsilon(x(t)) \sim N\left(0, \left(\frac{1+x(t)^2}{4}\right)^2\right)$
	\item Beta: $\epsilon(x(t)) \sim Beta(1, 11-10x(t))$
	\item Mixed normal: $\epsilon(x(t))$ is simulated from a mixture of $N(-1,1)$ and  $N(1,1)$ with mixing probability $x(t)$.
\end{itemize}
The true quantile trends and an example simulated data set is show in \Fig{sim1}. One hundred datasets were generated of sizes 300, 500 and 1000. 

\begin{figure}
	\includegraphics[width=\linewidth]{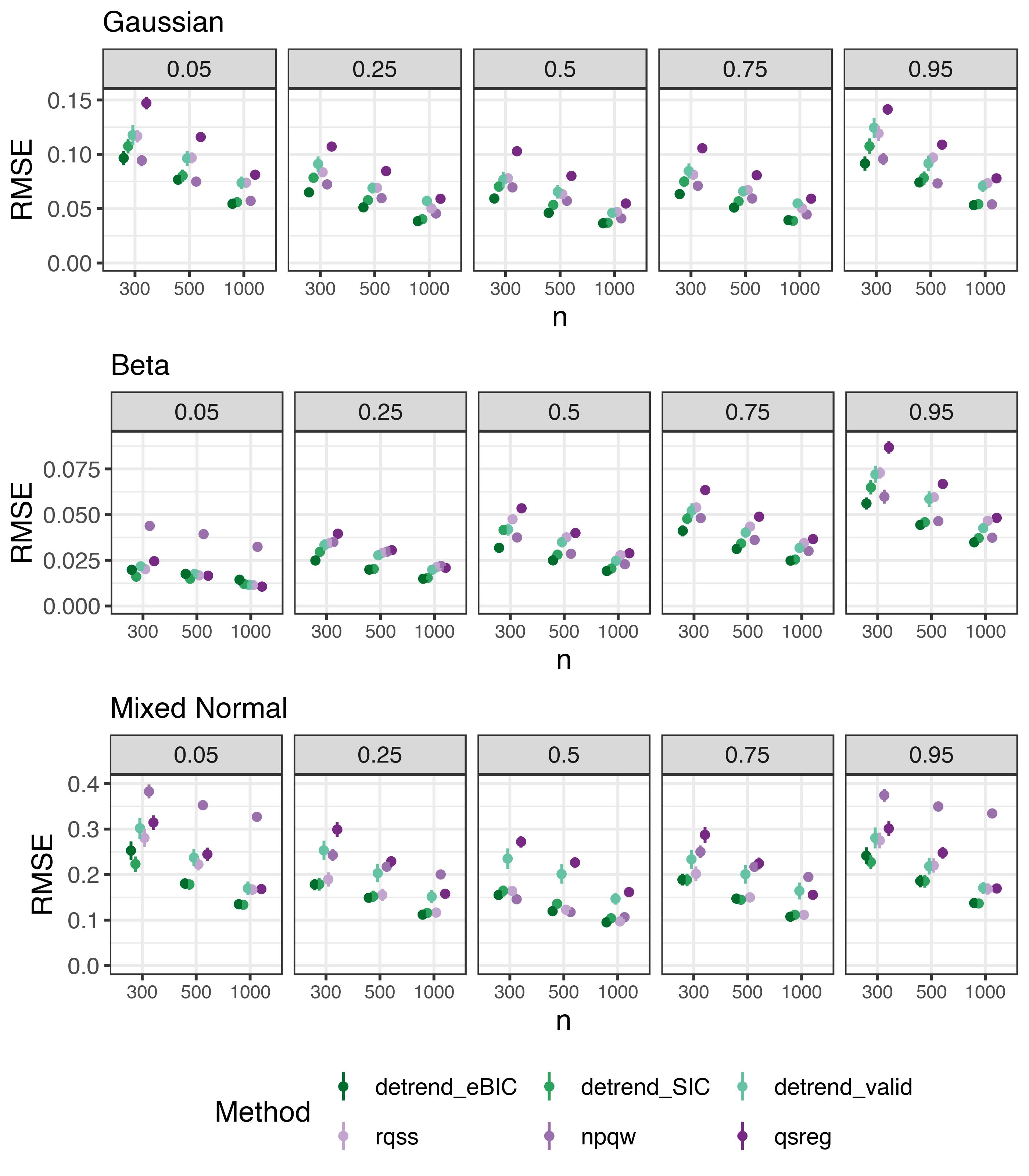}
	\caption{RMSE by design, method, quantile and data size. Points and error bars represent mean RMSE $\pm$ twice the standard error.}
	\label{fig:quantile_mse}
\end{figure}

Quantile trends were estimated for $\tau = \{0.05, 0.25, 0.5, 0.75, 0.95\}$ and the root mean squared error was calculated as $\mbox{RMSE}(\tau_j) = \sqrt{\frac{1}{n}\sum_{i=1}^n (\hat{\theta}_{ij} - \theta_{ij})^2}$, where $\theta_{ij}$ is the true value of the $\tau_j$th quantile of $y$ at $t=i$. \Fig{quantile_mse} shows the mean RMSE plus or minus twice the standard error for each method, quantile level, and sample size. In all three designs the proposed \texttt{detrend} methods are either better than or comparable to existing methods. Overall the \texttt{detrend\_eBIC} performs best. In the mixed normal design, specifically, our methods have lower RMSEs for the 5\textsuperscript{th} and 95\textsuperscript{th} quantiles. The \texttt{npqw} method performs particularly poorly on the mixed normal design due to the violation of the assumption that the data come from a scale-location model.

\subsection{Peak Detection}
\label{sec:peak_detection}
\begin{figure}[h!]
	\includegraphics[width = \linewidth]{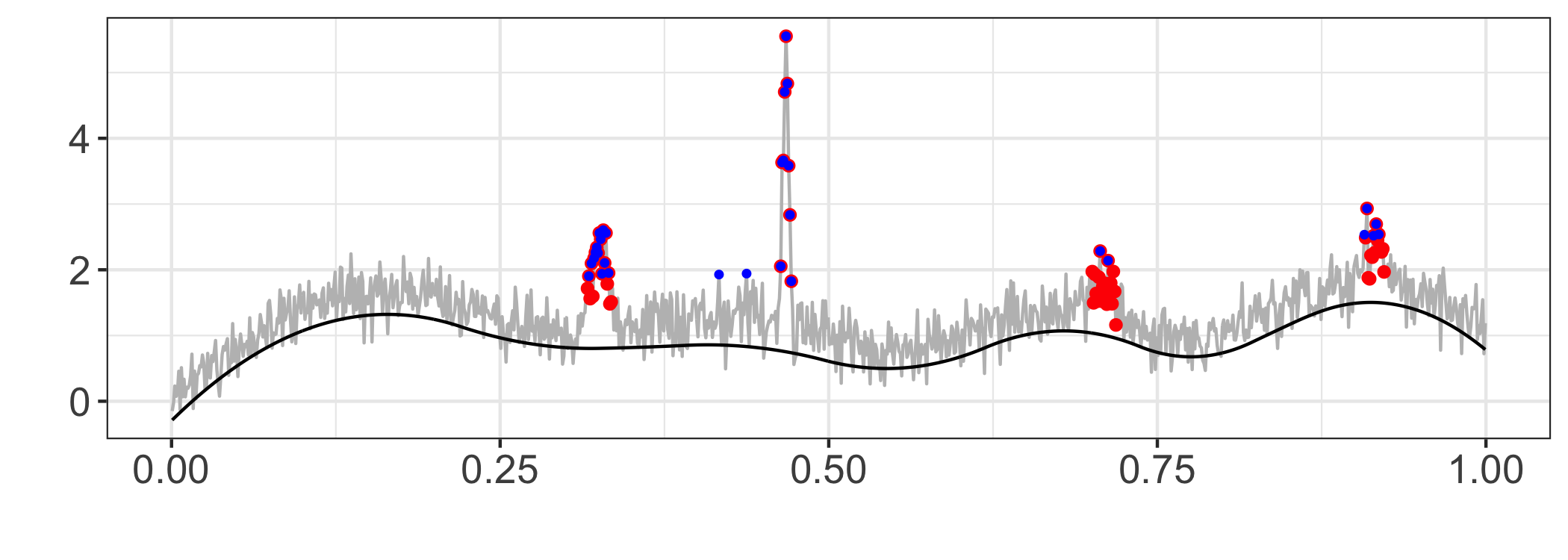}
	\caption{Example signal classification using threshold. Red indicates true signal $(\VE{y}{i} - \VE{\theta}{i} > 0.5)$, blue indicates observations classified as signal, i.e. values greater than 1.2 after baseline removal using \texttt{detrend\_eBIC}.}
	\label{fig:peaks_class_eg}
\end{figure}

The second simulation design is closely motivated by our air quality analysis problem. We assume that the measured data can be represented by
\begin{eqnarray*}
	y(t_i) & = & \theta(t_i) + s(t_i) + \VE{\varepsilon}{i},
\end{eqnarray*}
where $t_i = i$ for $i = 1, ..., n$, $\theta(t)$ is the drift component that varies smoothly over time, $s(t)$ is the true signal at time $t$, and $\VE{\varepsilon}{i}$ are i.i.d.\@ errors distributed as $N(0, 0.25^2)$. We generate $\theta(t)$ using cubic natural spline basis functions with degrees of freedom sampled from a Poisson distribution with mean parameter equal to $n/100$,  and coefficients drawn from an exponential distribution with rate 1. The true signal function $s(t)$ is assumed to be zero with peaks generated using the Gaussian density function. The number of peaks is sampled from a binomial distribution with size equal to $n$ and probability equal to $0.005$ with location parameters uniformly distributed between $1$ and $n-1$ and bandwidths uniformly distributed between $2$ and $12$. The simulated peaks were multiplied by a factor that was randomly drawn from a normal distribution with mean 20 and standard deviation of 4. An example dataset with 4 signal peaks is shown in \Fig{peaks_class_eg}. One hundred datasets were generated for each $n\in \{500, 1000, 2000, 4000\}$. 

\begin{figure}
	\includegraphics[width = \linewidth]{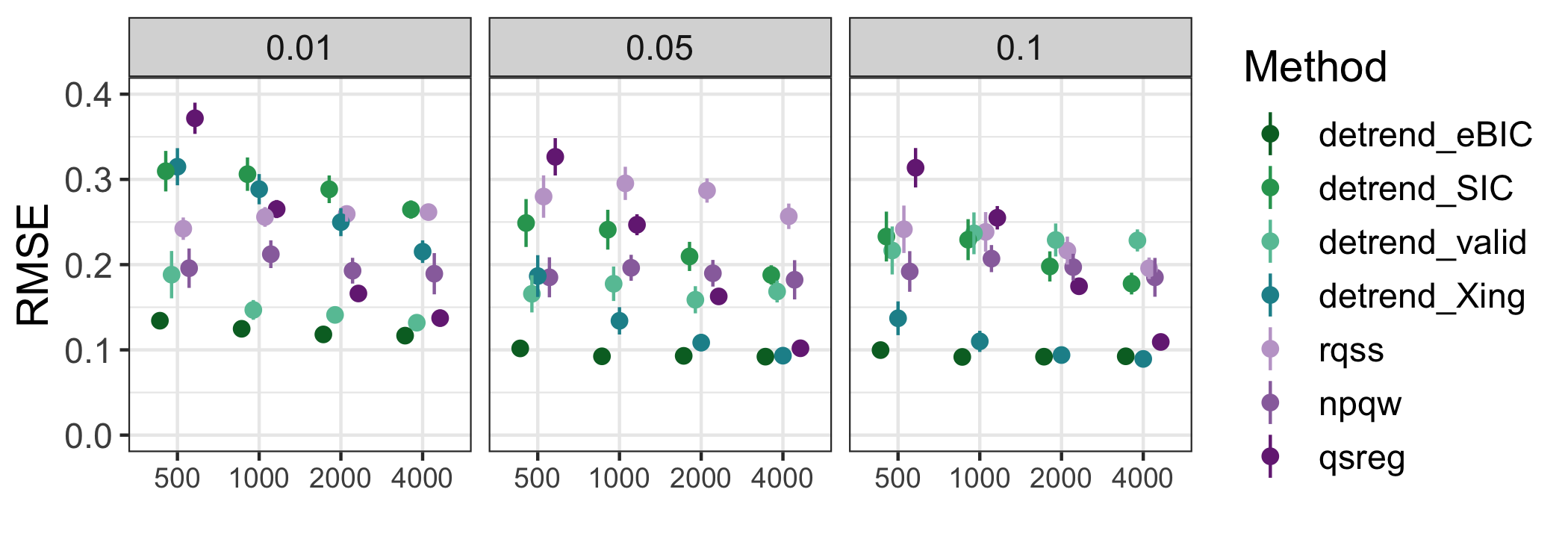}
	\caption{RMSE by method, quantile, and data size for peaks design.}
	\label{fig:peaks_rmse}
\end{figure}

We compare the ability of the methods to estimate the true quantiles of $y(t_i)-s(t_i)$  for $\tau \in \{0.01, 0.05, 0.1\}$ and calculate the RMSE (\Fig{peaks_rmse}). In this simulation study, our proposed method \texttt{detrend\_eBIC} method substantially outperforms the others. The \texttt{detrend\_Xing} method, which is the \texttt{detrend\_eBIC} method fit without the non-crossing constraints, performs similarly for larger quantiles and larger datasets. However, \texttt{detrend\_Xing} produces significantly worse estimates for more extreme quantiles ($\tau = 0.01$) and smaller data sets ($\tau = 0.05$ and $n = 500$). These results indicate that even when a single quantile is of interest, simultaneously fitting nearby quantiles and utilizing the non-crossing constraint can improve estimates when data is sparse by using information from nearby quantiles. The \texttt{qsreg} method is comparable to the \texttt{detrend\_eBIC} method on the larger datasets, but its performance deteriorates as the data size shrinks. The \texttt{npqw} and \texttt{rqss} methods both perform poorly on this design.

While minimizing RMSE is desirable in general, in our application, the primary metric of success is accurately classifying observations $\VE{y}{i}$ into signal present or absent. To evaluate the accuracy of our method compared to other methods we define true signal as any time point when the simulated peak value is greater than 0.5. We compare three different quantiles for the baseline estimation and four different thresholds for classifying the signal after subtracting the estimated baseline from the observations.  \Fig{peaks_class_eg} illustrates the observations classified as signal after subtracting the baseline trend compared to the ``true signal." 

To compare the resulting signal classifications, we calculate the class averaged accuracy (CAA), which is defined as
\begin{eqnarray*}
	\mbox{CAA} & = & \frac{1}{2}\left[\frac{\sum_{i=1}^n \One[\delta_i = 1 \cap \hat{\delta}_i = 1]}{\sum_{i=1}^n \One[\delta_i = 1]} + \frac{\sum_{i=1}^n \One[\delta_i = 0 \cap \hat{\delta}_i=0]}{\sum_{t=i=1}^n \One[\delta_i = 0]}\right].
\end{eqnarray*}
where $\delta_i \in \{0,1\}$ is the vector of true signal classifications and $\hat{\delta}_i \in \{0,1\}$ is the vector of estimated signal classifications, namely $\hat{\delta}_i = \One(\VE{y}{i} - \hat{\theta}_{i} > 0.5)$. We use this metric because our classes tend to be very imbalanced with many more zeros than ones. The CAA metric should give a score close to 0.5 both for random guessing and also for trivial classifiers such as $\hat{\delta}_i = 0$ for all $i$.

\begin{figure}[h!]
	\includegraphics[width = \linewidth]{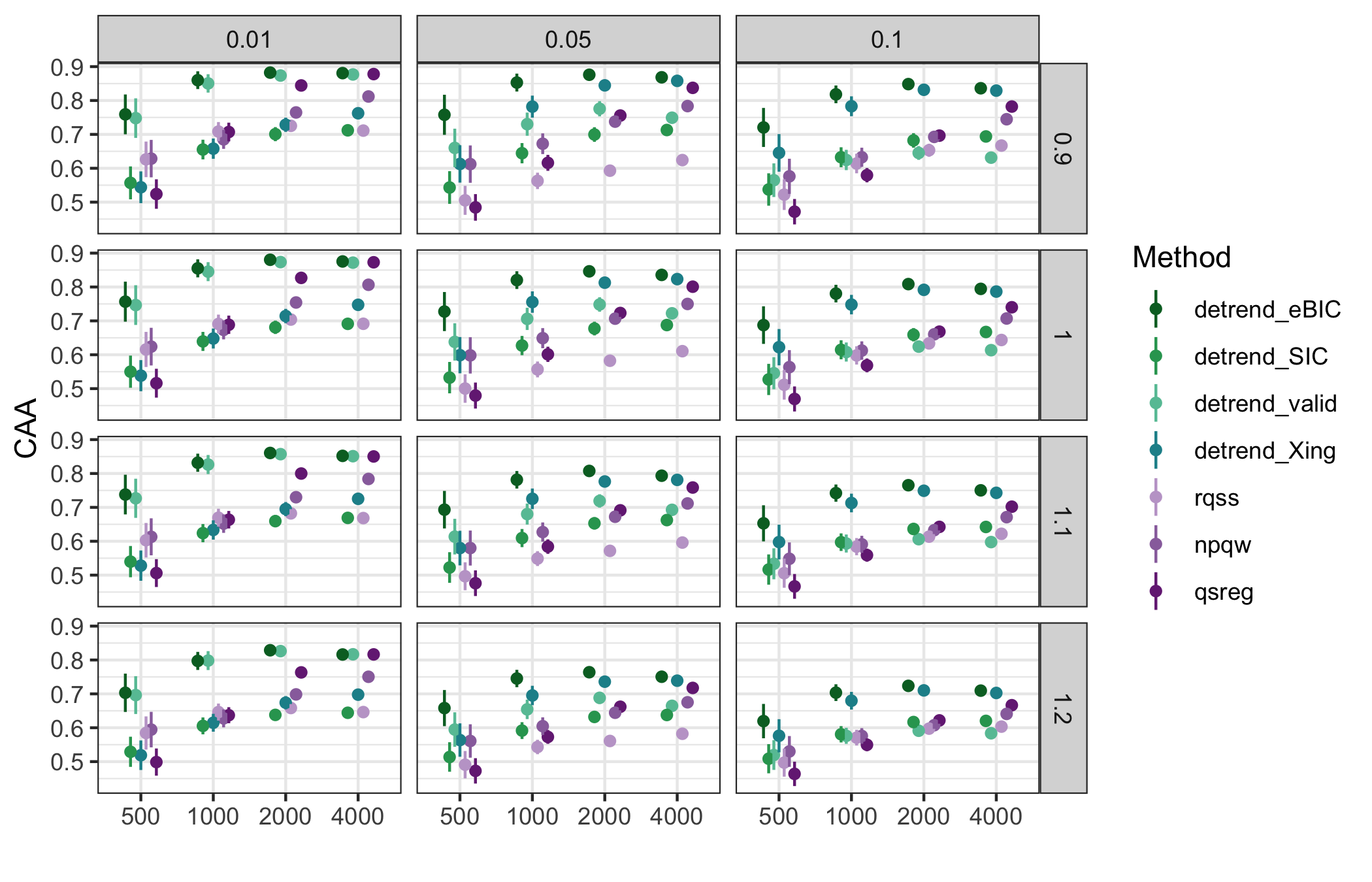}
	\caption{Class averaged accuracy by threshold, data size, and method (1 is best 0.5 is worst).}
	\label{fig:CAA}
\end{figure}

Our \texttt{detrend\_eBIC} method results in the largest CAA values (\Fig{CAA}) in addition to the smallest RMSE values (\Fig{peaks_rmse}). While \texttt{qsreg} was competitive with our method in some cases, in the majority of cases the largest CAA values for each threshold were produced using the \texttt{detrend\_eBIC} method with the 1\textsuperscript{st} or 5\textsuperscript{th} quantiles.

\FloatBarrier

\section{Analysis of Air Quality Data}

\label{sec:application}
The low-cost ``SPod" air quality sensors output a time series that includes a slowly varying baseline,
the sensor response to pollutants, and high frequency random noise. These sensors record measurements every second and are used to monitor pollutant concentrations at the perimeter of industrial facilities. Time points with high concentrations are identified and compared with concurrent wind direction and speed. Ideally, three co-located and time aligned sensors (as shown in \Fig{raw_spod}) responding to a pollutant plume would result in the same signal classification after baseline trend removal and proper threshold choice. We first illustrate the difference between our \texttt{detrend\_eBIC} method,  hereafter referred to as \texttt{detrendr}, and \texttt{qsreg} using data from 13:10 to 15:10 from \Fig{raw_spod} (\Sec{short-app}). We then compare the methods on the complete day shown in \Fig{raw_spod}, estimating trends by applying the \texttt{qsreg} method to 2 hour non-overlapping windows of the data and \Alg{admm} to the entire day. We focus on this day because data from three sensors was available. Finally, we examine an entire week of measurements from two co-located sensors (\Sec{long-app}). 

\subsection{Short series of air quality measurements}
\label{sec:short-app}
\begin{figure}
	\includegraphics[width = \linewidth]{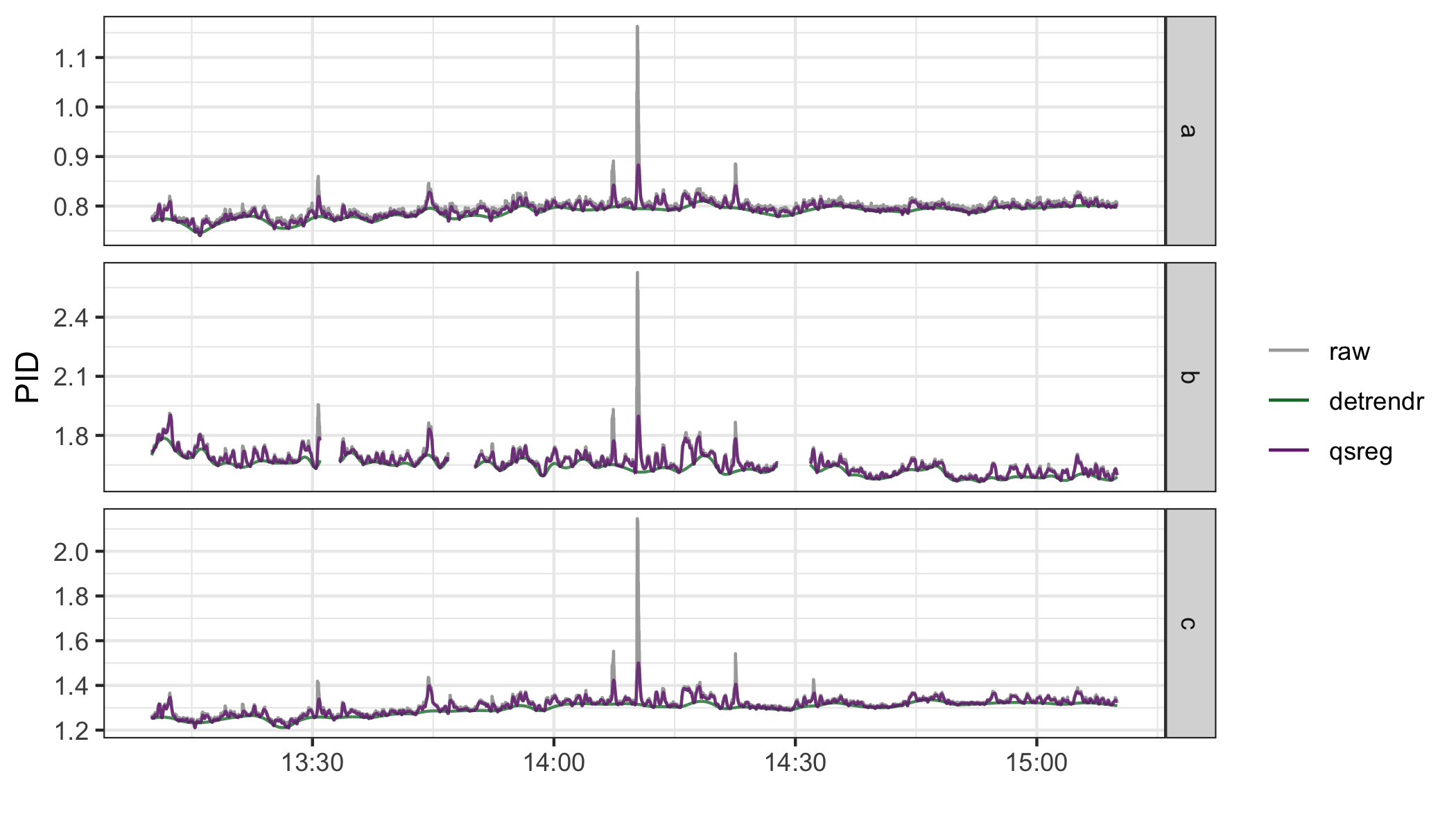}
	\caption{Estimated 5\textsuperscript{th} quantile trends for SPods a, b, and c, using \texttt{qsreg} and \texttt{detrendr}. SPod c contains some missing values that were interpolated before the trends were estimated. \texttt{qsreg} is more influenced by the signal component apparent on both nodes resulting in under-smoothing.}
	\label{fig:short-trends}
\end{figure}

We compare our \texttt{detrendr} method with the \texttt{qsreg} method on a two-hour subset of one-second  SPod data (n=7200) both to facilitate visualization and because the \texttt{qsreg} method cannot handle all 24 hours simultaneously. We estimate the baseline trend using $\tau = \{0.01, 0.05, 0.1\}$ and compare three thresholds for classifying the signal. The thresholds are calculated as the 90\textsuperscript{th}, 95\textsuperscript{th}, and 99\textsuperscript{th} quantiles of the de-trended series for each SPod. If there is signal present in the dataset, values above these thresholds should occur simultaneously on all three SPods. We do not use class-averaged accuracy to compare the signal classifications because we do not have a reference value to define as the ``true" signal. Instead, we compute the variation of information (VI) which compares the similarity between two classifications. Given the signal classifications for SPods a and b, $\delta^{(a)}_i \in \{0,1\}$ and $\delta^{(b)}_i\in\{0,1\}$, for $i \in \{1, ..., n\}$ the VI is defined as:
\begin{eqnarray*}
	r_{jk} & = & \frac{1}{n}\sum_i \One\left(\delta^{(a)}_i = j  \cap \delta^{(b)}_i = k\right)\\
	\VI(a, b) & = & -\sum_{j,k} r_{jk} \left[ \log \left(\frac{r_{jk}}{\frac{1}{n}\sum_i \One(\delta^{(a)}_i = j)}\right) +
	\log \left(\frac{r_{jk}}{\frac{1}{n}\sum_i \One(\delta^{(b)}_i = k)}\right) \right]
\end{eqnarray*}
where $(j,k) \in \{(0,0), (0,1), (1,0), (1,1)\}$. The $\VI$ is a distance metric for measuring similarity of classifications and will be 0 if the classifications are identical and increase as the classifications become more different.

\begin{figure}
	\includegraphics[width = \linewidth]{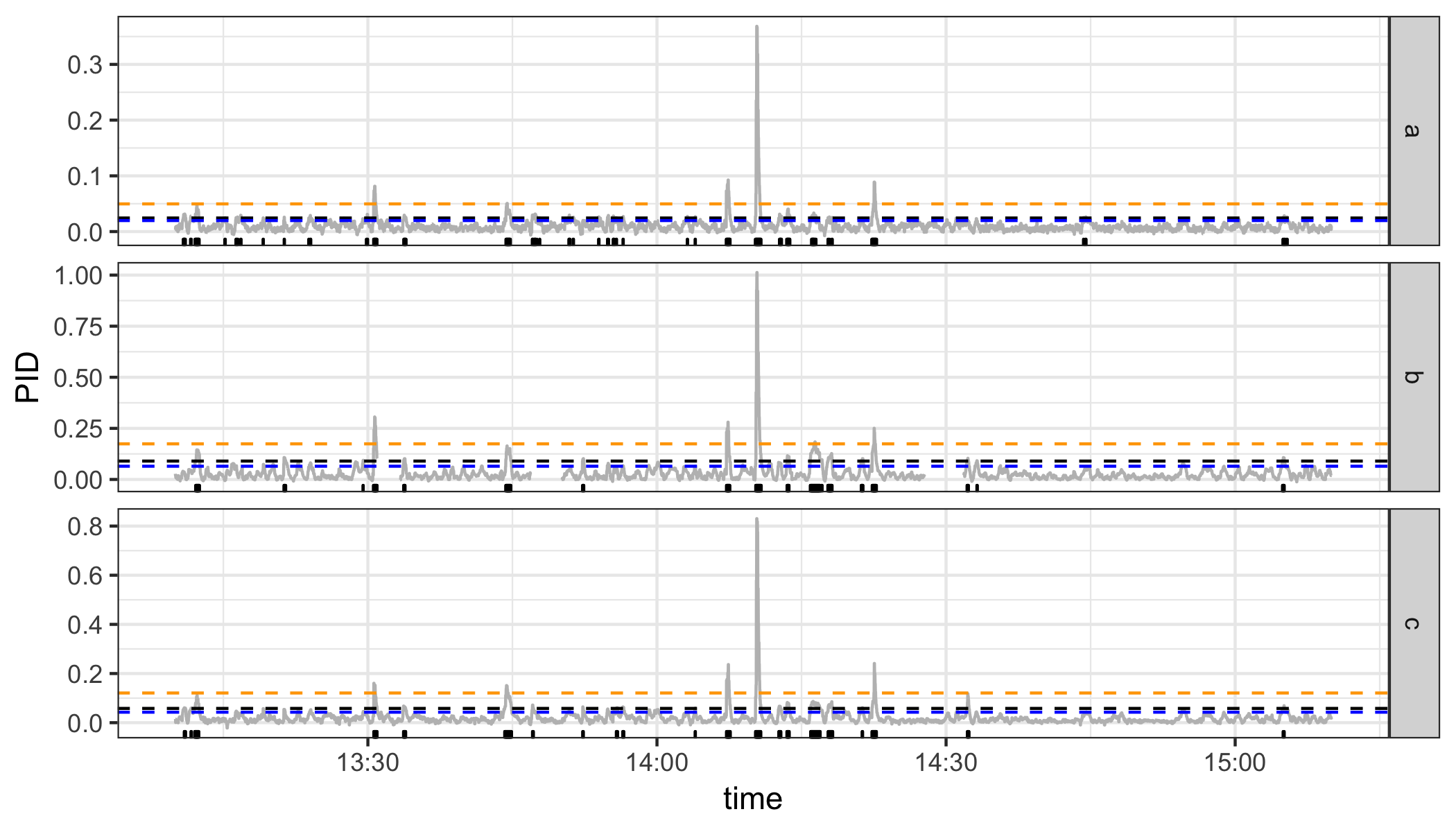}
	\caption{Rugplot showing locations of signal after baseline removal using the \texttt{detrendr} estimate of 5th quantile. Horizontal dashed lines represent the thresholds calculated using the 90\textsuperscript{th}, 95\textsuperscript{th}, and 99\textsuperscript{th} quantiles of the detrended data. The 95\textsuperscript{th} quantile (black) was used to classify the signal shown as vertical lines at the bottom of the plot.} 
	\label{fig:rugplot}
\end{figure}

\begin{figure}
	\centering
	\includegraphics[width = 1\linewidth]{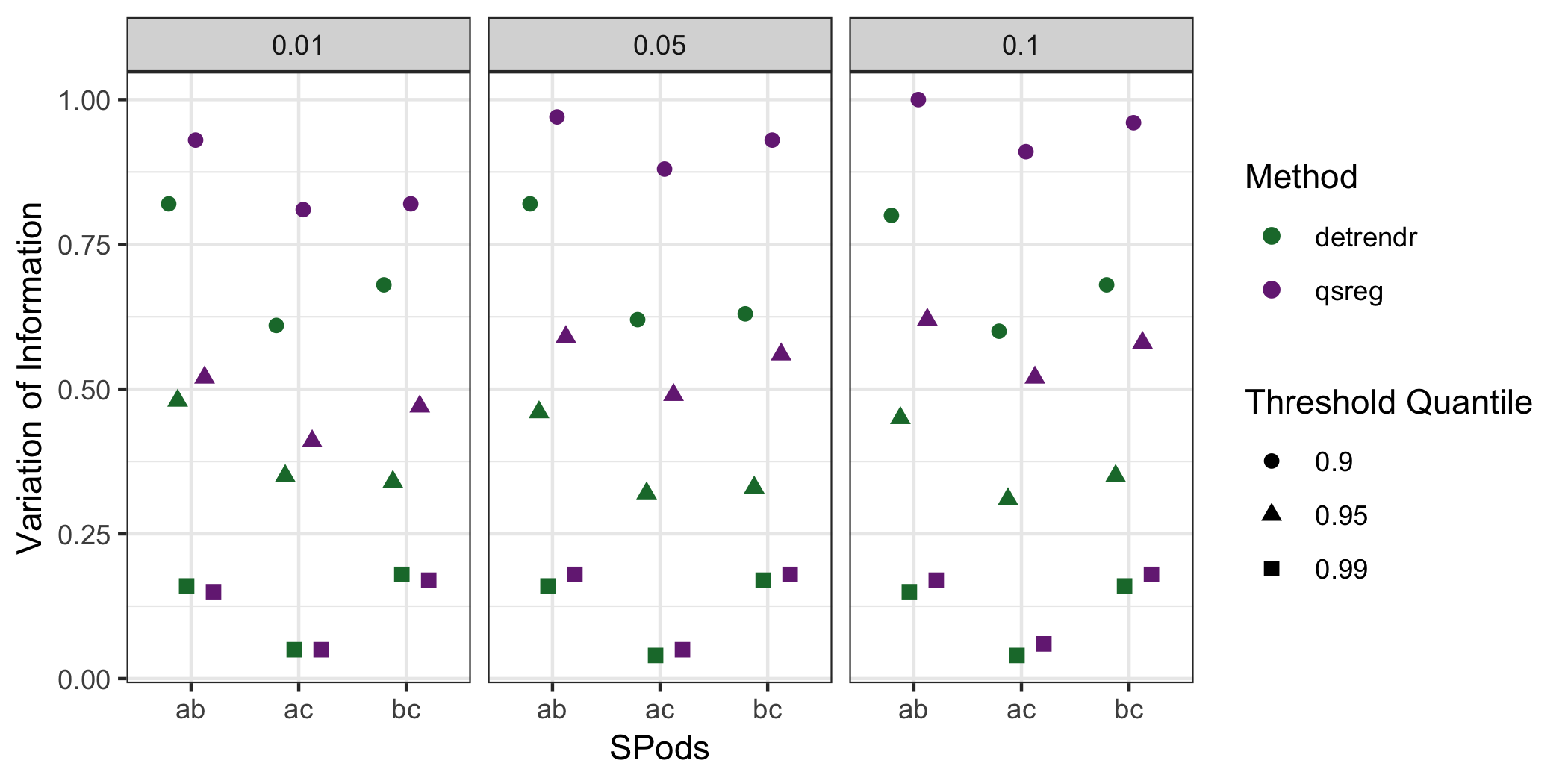}
	\caption{Variation of Information between sensors by method (color), quantile (columns) and threshold (shape) for two hour time period. }
	\label{fig:vi}
\end{figure}

\Fig{short-trends} shows the estimated 5\textsuperscript{th} quantile trends from each method for each SPod. The \texttt{detrendr} method results in a smoother baseline estimate while the \texttt{qsreg} method absorbs more of the peaks obscuring some of the signal.  \Fig{rugplot} shows the series after subtracting the \texttt{detrendr} estimate of the 5\textsuperscript{th} quantile and classifying the signal using the 95\textsuperscript{th} quantile of the detrended data. The 90\textsuperscript{th} and 99\textsuperscript{th} thresholds are also shown for comparison in blue and orange, respectively. The largest peaks at 14:10 are easily identified as signal, but good baseline estimates also enable proper classification of the smaller peaks like the one at 15:12. The under-smoothing of the \texttt{qsreg} method results in less similar signal classifications and higher VI values for the 90\textsuperscript{th} and 95\textsuperscript{th} quantile thresholds (\Fig{vi}). However, when the 99\textsuperscript{th} threshold is used only the highest observations are classified as signal and the baseline estimation method isn't as important (\Fig{vi}).  

\subsection{Long series of air quality measurements}
\label{sec:long-app}
\begin{figure}
	\centering
	\includegraphics[width = 1\linewidth]{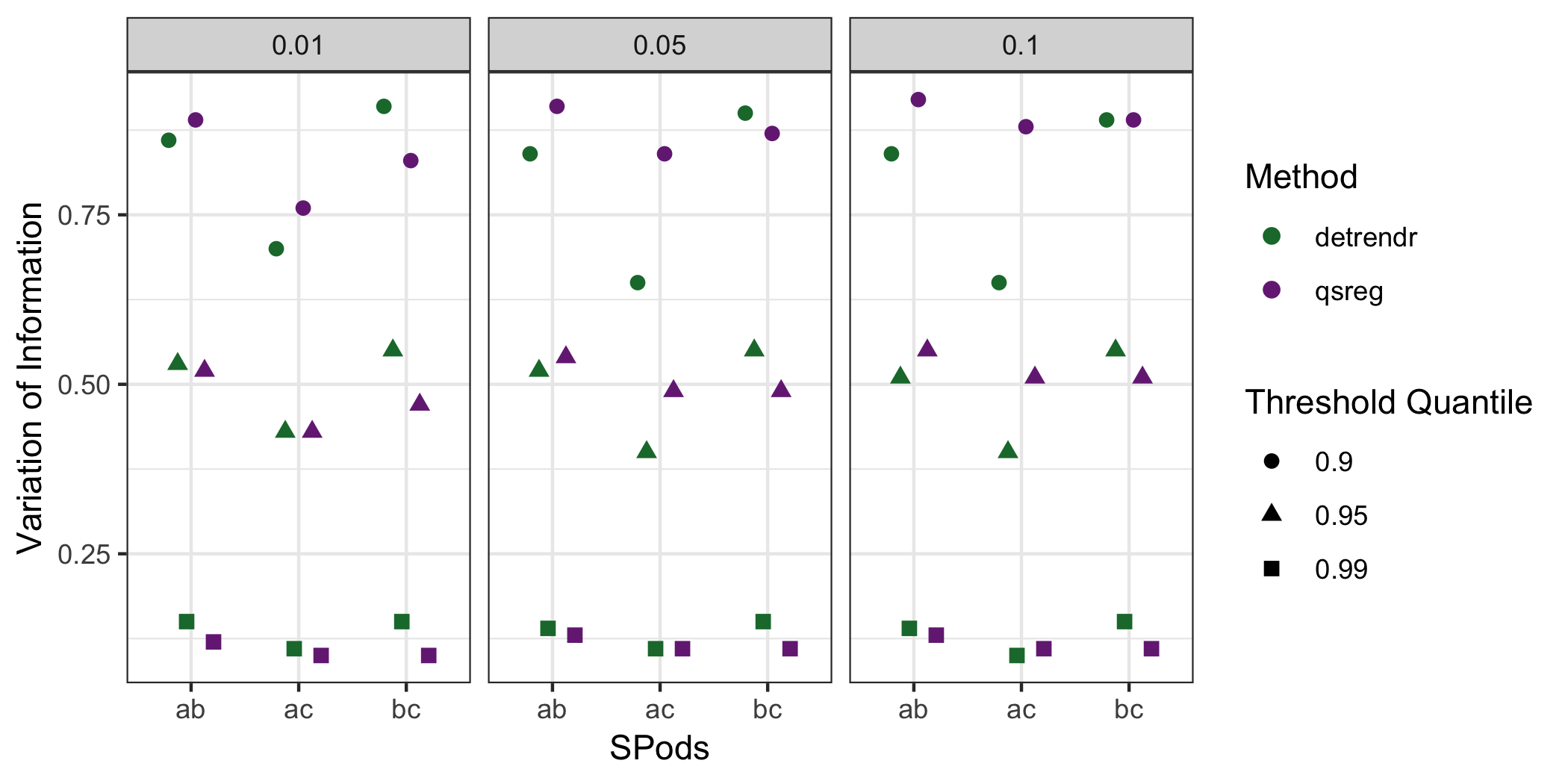}
	\caption{Variation of Information between sensors by method (color), quantile (columns) and threshold (shape) for full day.}
	\label{fig:vi_day}
\end{figure}

\begin{figure}
	\centering
	\includegraphics[width = 0.3\linewidth]{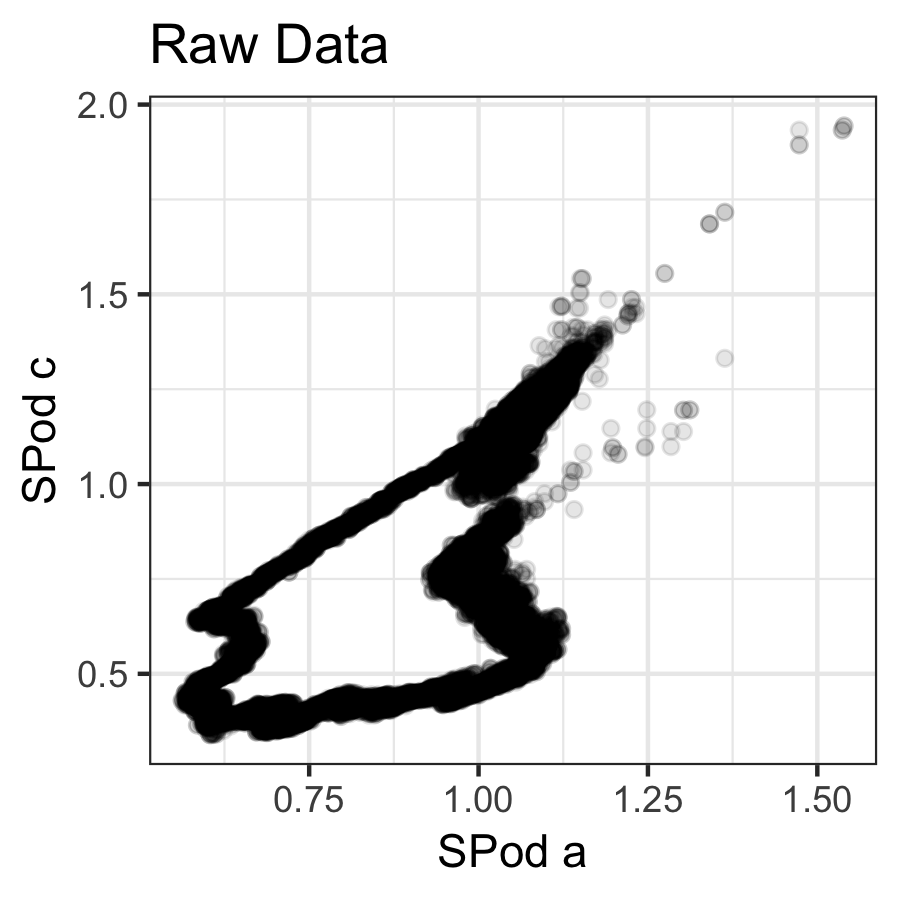}
	\includegraphics[width = 0.3\linewidth]{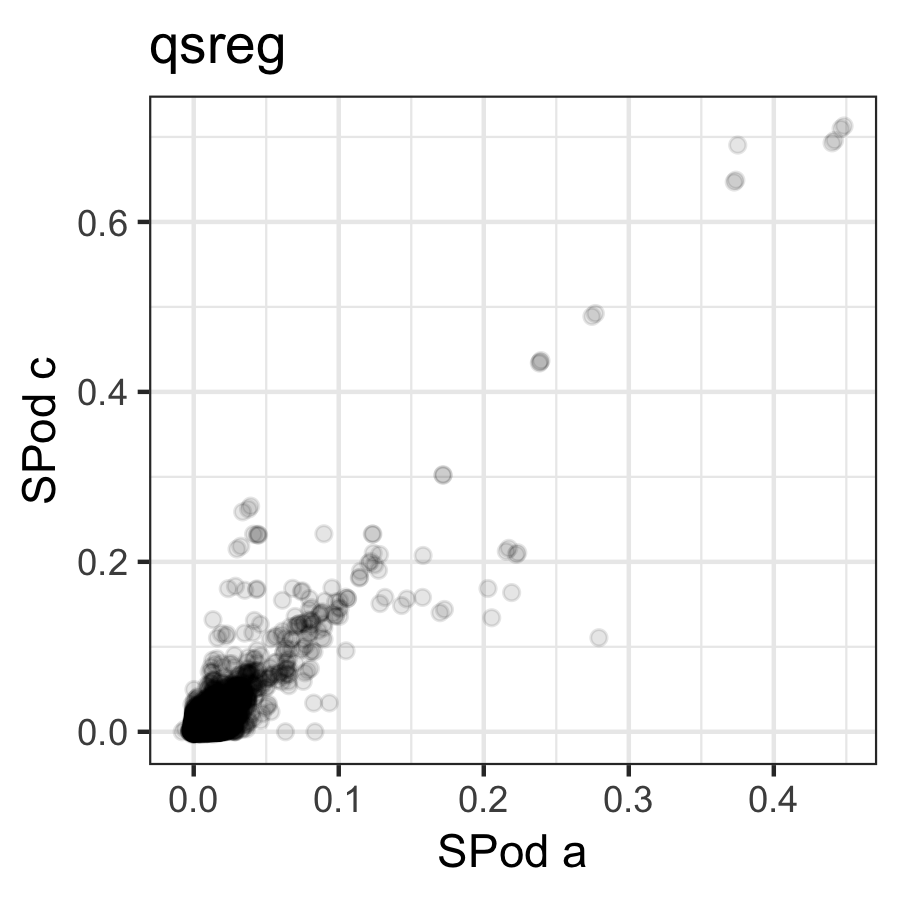}
	\includegraphics[width = 0.3\linewidth]{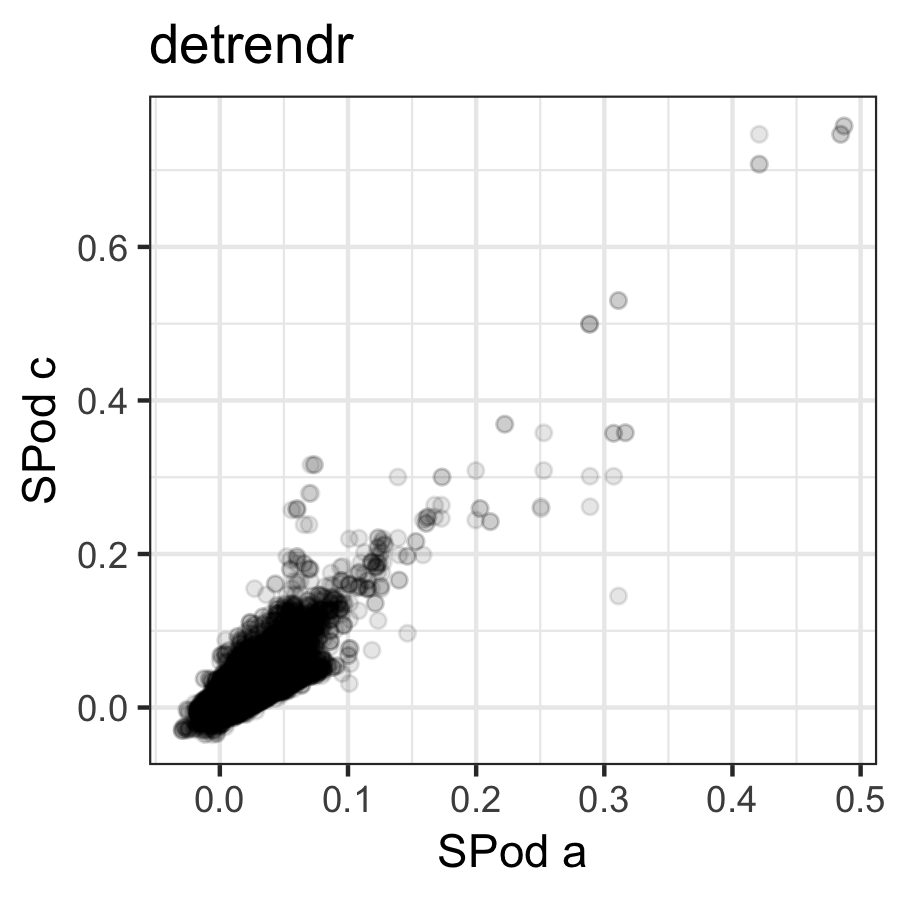}
	\caption{SPod a versus SPod c before and after de-trending with \texttt{qsreg} and \texttt{detrendr}. }
	\label{fig:scatter_day}
\end{figure}

\Alg{admm} was used to remove the baseline drift from the full day of data  (\Fig{raw_spod}) consisting of 86,400 observations per SPod and compared to the series detrended using the \texttt{qsreg} trends estimated using non-overlapping 2 hour windows. As in the shorter illustration, the \texttt{detrendr} method results in generally lower VI scores than the \texttt{qsreg} method (\Fig{vi_day}). The \texttt{detrendr} method also results in better correlation in the detrended series as is illustrated in (\Fig{scatter_day}). The Spearman correlation coefficients for SPods a and b, SPods a and c, and SPods b and c after removing the 5\textsuperscript{th} quantile trend using \texttt{detrendr} were 0.37, 0.75, and 0.43, compared with 0.07, 0.24, and 0.16 using \texttt{qsreg}. The noise variance was higher for SPod b than for SPods a and c resulting in lower correlation and higher VI values for the ab and bc metrics compared with ac. 

\begin{figure}
	\includegraphics[width = 0.6\linewidth]{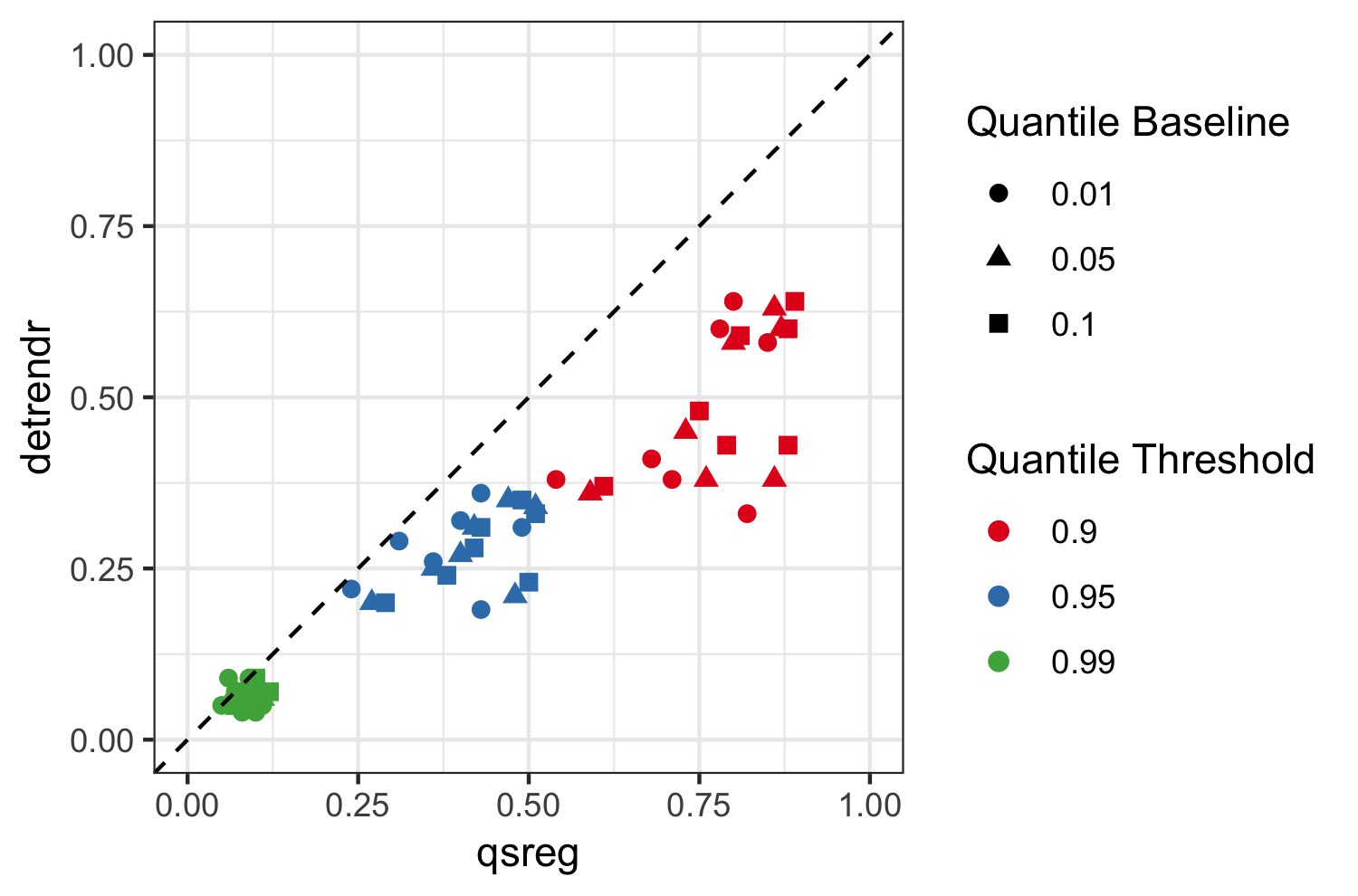}
	\caption{Variation of Information (VI) for \texttt{detrendr} and \texttt{qsreg} by quantile trend and threshold. Each point represents a full day of data. The dashed line represents y=x. In most cases \texttt{detrendr} results in a lower VI than \texttt{qsreg}.}
	\label{fig:vi_week}
\end{figure}

Finally, we estimated the quantile trends for 7 days of measurements of two co-located SPods. \Fig{vi_week} demonstrates the improvement in classification similarity when using \texttt{detrendr}. Each point represents a day of measurements and all points that fall below the dashed line have more similar classifications using \texttt{detrendr} compared to \texttt{qsreg}. The improvement of \texttt{detrendr} over \texttt{qsreg} is more severe at lower thresholds. This indicates that \texttt{detrendr} gives greater agreement on signal classification when the methods are tuned to deliver positive classifications more frequently. 

\section{Conclusion and Discussion}
\label{sec:discussion}

We have expanded the quantile trend filtering method by implementing a non-crossing constraint, a new algorithm for processing large series, and proposing a modified criteria for smoothing parameter selection. Furthermore, we have demonstrated the utility of quantile trend filtering in both simulations and applied settings. Our ADMM algorithm for large series both reduces the computing time and allows trends to be estimated on series that cannot be estimated simultaneously while our scaled extended BIC criterion was shown to provide better estimates of quantile trends in series with and without a signal component. 

In our application to low cost air quality sensor data, we have shown that the baseline drift in low cost air quality sensors can be removed through estimating quantile trends, but the data size was too large for existing methods to be computationally feasible. While \texttt{qsreg} cannot feasibly handle more than a few hour windows of data, our new methods were able to process 24 hours simultaneously and deliver signal classifications that were more consistent between the two sensors for a week of data (168 hours). 

In the future, quantile trend filtering could be extended to observations measured at non-uniform spacing by incorporating the distance in covariate spacing into the differencing matrix. It could also be extended to estimate smooth spatial trends by a similar adjustment to the differencing matrix based on spatial distances between observations.

\section{Acknowledgments}
The authors greatly appreciate Eben Thoma at the US Environmental Protection Agency for providing the SPod datasets. The authors are also thankful for the Oak Ridge Institute of Science Fellowship that partially supported this work. Plots were created using ggplot2 \citep{ggplot2}.  


\bigskip
\begin{center}
	{\large\bf SUPPLEMENTARY MATERIAL}
\end{center}

\begin{description}
	
	\item[R-package for detrend routine:] R-package detrendr containing code to perform the methods described in the article. (GNU zipped tar file also available at https://github.com/halleybrantley/detrendr)
	
\end{description}

\bibliographystyle{imsart-nameyear}
\bibliography{detrendify}

\end{document}